%% file: Festschrift_Reif.tex
\documentclass[runningheads]{llncs}
\usepackage[T1]{fontenc}
\usepackage{graphicx}
\usepackage{hyperref}
\usepackage{color}

\usepackage{varioref}
\labelformat{section}{Section~#1}
\labelformat{subsection}{Section~#1}
\labelformat{subsubsection}{Section~#1}
\labelformat{figure}{Figure~#1}
\labelformat{table}{Table~#1}
\labelformat{equation}{Equation~#1}

\usepackage[utf8]{inputenc}
\usepackage[english]{babel}

\usepackage{csquotes}
\usepackage{caption}
\usepackage{wrapfig}
\usepackage{booktabs}
\usepackage{array}
\usepackage{amssymb}
\usepackage[cmex10]{amsmath}
\usepackage{algorithm}
\usepackage{algpseudocode}
\usepackage{xcolor}
\usepackage{listings}
\lstdefinestyle{IDL}{
	backgroundcolor=\color{lime},
	rulecolor=\color{lightgray},
	tabsize=2,
	basicstyle=\ttfamily\footnotesize,
	otherkeywords={DEFAULT_VELOCITY},
	morekeywords={DEFAULT_VELOCITY},
	keywordstyle=\color{magenta},
	frame=single,
	framextopmargin=5pt,
	framexbottommargin=5pt,
	showstringspaces=false,
	showtabs=false,
	numbers=none,
	breakatwhitespace=true,
	captionpos=b, 
	belowcaptionskip=1\baselineskip 
}
\usepackage{pgfplots}
\usepackage{tikz}
	\usetikzlibrary{arrows,positioning,backgrounds,fit,trees} 
	\usetikzlibrary{fadings,shapes.geometric}
	\usetikzlibrary{decorations,scopes,calc,decorations.pathreplacing, decorations.markings}
	\usetikzlibrary{backgrounds}
\usepackage{tikz-3dplot}
\usepackage{adjustbox}

\newcommand{\cuboid}[4]{
    \draw[fill=gray] (#1+#4,#2-#4,#3-#4) -- (#1+#4,#2+#4,#3-#4) -- (#1+#4, #2+#4, #3+#4) -- (#1+#4, #2-#4, #3+#4) -- cycle;
    \draw[fill=gray] (#1+#4,#2+#4,#3-#4) -- (#1-#4,#2+#4,#3-#4) -- (#1-#4, #2+#4, #3+#4) -- (#1+#4, #2+#4, #3+#4) -- cycle;
    \draw[fill=gray] (#1+#4,#2+#4,#3+#4) -- (#1-#4,#2+#4,#3+#4) -- (#1-#4, #2-#4, #3+#4) -- (#1+#4, #2-#4, #3+#4) -- cycle;
}

\usepackage{todonotes}

\usepackage{environ}

\newif\ifshowtemplate
\showtemplatetrue

\NewEnviron{template}[4]{
	\ifshowtemplate
		\color{blue}
		\textit{Hints: #1} \\
		\textit{Limit: #2} \\ 
		\textit{Author: #3}\\ 
		\textit{Remaining tasks: #4} \ \\
		\BODY
	\fi
}

\makeatletter
\tikzoption{canvas is plane}[]{\@setOxy#1}
\def\@setOxy O(#1,#2,#3)x(#4,#5,#6)y(#7,#8,#9)%
{\def\tikz@plane@origin{\pgfpointxyz{#1}{#2}{#3}}%
	\def\tikz@plane@x{\pgfpointxyz{#4}{#5}{#6}}%
	\def\tikz@plane@y{\pgfpointxyz{#7}{#8}{#9}}%
	\tikz@canvas@is@plane
}
\makeatother

\begin{document}
\graphicspath{{img/}}
\title{How to Drawjectory? - Trajectory Planning using Programming by Demonstration}
\titlerunning{How to Drawjectory?}
\author{Leonhard Alkewitz\inst{1}\orcidID{0009-0002-5693-4526} \and
Timo Zuccarello\inst{1}\orcidID{0009-0003-0992-8155} \and 
Alexander Raschke\inst{1}\orcidID{0000-0002-6088-8393} \and
Matthias Tichy\inst{1}\orcidID{0000-0002-9067-3748}}
\authorrunning{L. Alkewitz et al.}
\institute{Institute of Software Engineering and Programming Languages, Ulm University, 89081 Ulm, Germany
\email{\{firstname.lastname\}@uni-ulm.de}}
\maketitle              %
\begin{abstract}
	A flight trajectory defines how exactly a quadrocopter moves in the three-dimensional space from one position to another. 
	Automatic flight trajectory planning faces challenges such as high computational effort and a lack of precision. 
	Hence, when low computational effort or precise control is required, programming the flight route trajectory manually might be preferable. However, this
	requires in-depth knowledge of how to accurately plan flight trajectories in three-dimensional space.

	We propose planning quadrocopter flight trajectories manually using the \textit{Programming by Demonstration (PbD)} approach -- simply drawing the trajectory in the three-dimensional space by hand. This simplifies the planning process and reduces the level of in-depth knowledge required. 
	We implemented the approach in the context of the \textit{Quadcopter Lab} at Ulm University.

	In order to evaluate our approach, we compare the 
	precision and accuracy of the trajectories drawn by a user using our approach as well as the required time with those manually programmed using a domain specific language.
	The evaluation shows that the \textit{Drawjectory} workflow is, on average, $78.7$ seconds faster without a significant loss of precision, 
	shown by an average deviation $6.67$ cm.

\end{abstract}

\input{sections/introduction}

\input{sections/personal_reference}

\input{sections/foundatations}

\input{sections/drawjectory_workflow}

\input{sections/evaluation}

\input{sections/related_work}

\input{sections/conclusion}

\bibliographystyle{splncs04}
\bibliography{literature/references}
\end{document}

%% file: sections/introduction.tex
\section{Introduction}
\label{sec:introduction}

\begin{quote}\em
    Life is full surprising coincidences. Wolfang Reif twice had -- in totally unrelated circumstances -- a profound impact on the Institute of Software Engineering and Programming Languages at Ulm University. 

    First, Wolfgang Reif was professor at the institute -- then called Software Engineering and Compiler Construction -- from 1994 to 2000 (before moving on to heading the Institute of Software and Systems Engineering at the University of Augsburg) working on formal verification and the KIV system. He put Ulm on the international map as a renowned place for formal methods research in the software engineering community.

    Second, Wolfgang Reif gave me (Matthias Tichy) in 2010 the first opportunity for independent research and teaching by offering me the position as acting professor for Self-Organizing Systems at the University of Augsburg for one year. While my goal was then to leave academia to join industry, seizing that opportunity allowed me to grow as a person and as an academic and put me on my academic path. 

    Finally, those two unrelated circumstances surprisingly met when i was offered to become head of the Institute of Software Engineering and Compiler Construction at Ulm University in 2015. Impressed by the research of Wolfgang Reif's group in the area of autonomous quadrocopters, i decided to build a quadrocopter lab as a research pillar in Ulm as well. Wolfgang Reif and his group, particularly, Andreas Angerer and Alwin Hoffmann, have been a tremendous help in building up that lab and our research. Hence, we chose to present one of our lab's recent outcomes in this paper to commemorate Wolfgang Reif's 65th birthday.   
\end{quote}

Quadrocopters have a wide range of applications, from automated parcel delivery to search and rescue operations.
Therefore, it is not surprising that the topic of planning flight routes or trajectories for drones is becoming increasingly important \cite{Herdel2022beyondscopingreview,Aggarwal2020Pathplanningtechniques}.

Generally speaking, trajectory planning \textquote{consists in assigning a time law to the geometric path} \cite{Gasparetto2015Pathplanningtrajectory}.
Trajectory planning distinguishes two planning strategies: manual and automatic planning.
When considering automatic trajectory planning, there are several promising approaches, such as
genetic algorithms, artificial neural networks, or simple A* algorithms \cite{Aggarwal2020Pathplanningtechniques,Gasparetto2015Pathplanningtrajectory}.

However, some tasks require high precision and specific points of interest to be visited. Thus, the mentioned strategies may not always be ideal for trajectory planning \cite{Xu2019Robottrajectorytracking}.
For instance, an old building in danger of collapse might need to be screened by a drone with the drone having to check certain rooms.
In this case, the creation of a route through the building, encompassing all points of interest, can be a particularly time-consuming and error-prone endeavor, 
particularly when the objective is to implement an automatic planning process, as this approach does not permit any form of intervention in the selection process of the points to be visited. 

One possible solution to ensure that all points of interest are visited is manual trajectory planning, whereby the drone's flight path is not determined automatically but by hand.
There are alternative approaches to manual planning including 
i) programming the path (mostly using low-code approaches) and
ii) creating interactive points in a 3D environment on a PC \cite{Witte2019HybridEditorFast,Hoppe2019Dronosflexibleopen}.
However, these methods often require knowledge of either how to program or how to interact with the points, 
making planning unintuitive.

Accordingly, there is a desire for a user-friendly and still accurate method for manual trajectory planning,
such as demonstrating the trajectory in the real world.
This concept of \textquote{transfer[ring] new skills to a machine by relying on demonstrations from a user} \cite{Calinon2018Learningdemonstrationprogramming} is called \textit{Programming by Demonstration}.
Due to its high intuitiveness, it is ideal for domain experts in a certain topic who need the support of quadrocopters, despite their potential lack of experience in controlling quadrocopters \cite{Melchior2012Graphbasedtrajectory},
which is why programming by demonstration has seen significant growth \cite{Ravichandar2020Recentadvancesrobot}.

Based on the aforementioned problem and goals, we formulate the following research questions:

\begin{description}
	\item[RQ 1] How to automatically plan a trajectory by demonstrating the flight path once?
	\item[RQ 2] How does trajectory planning by demonstration compare against trajectory planning by manual programming using a domain specific language in terms of accuracy and effort?
\end{description}

We developed an approach for planning trajectories from a user's demonstration, thus applying the
\textit{programming by demonstration} paradigm answering RQ1. Specifically, we track
the demonstration, i.e., the point sequence of the desired flight path. Subsequently, a selection of points,
designated as waypoints, is made from the recorded point sequence. These waypoints are then used to interpolate a trajectory using natural cubic splines.

To answer RQ2, we evaluate the accuracy and the planning effort by systematically using both approaches to plan three different classes of trajectories: planar geometric figures, planar non-trivial figures, 3D figures. We measure the time to plan those trajectories as well as the accuracy of the planned trajectory compared to the intended trajectory. The evaluation shows that the \textit{Drawjectory} workflow is, on average, $78.7$ seconds faster without a significant loss of precision, 
shown by an average deviation $6.67$ cm compared to using a domain specific language to manually program the trajectories.

\ref{sec:quadcopter_lab} and \ref{sec:foundatations} introduces the quadrocopter lab setting as well as relevant foundations of trajectory planning.
Thereafter, \ref{sec:drawjectory_workflow} presents our trajectory planning by demonstration approach and the proof-of-concept implementation. 
\ref{sec:evaluation} presents the results of the evaluation. After discussing related work in\ref{sec:related_work}, we conclude in  \ref{sec:conclusion} by summarizing the work and addressing limitations as well as discussing potential future work.

%% file: sections/personal_reference.tex
\section{Overview of \textit{Quadcopter Lab}}
\label{sec:quadcopter_lab}

The proposed \textit{Drawjectory} workflow and its proof-of-concept implementation are incorporated into the 
quadrocopter laboratory (\textit{Quadcopter Lab}) at the Institute of Software Engineering and Programming Languages at Ulm University.
The quadrocopter laboratory is a spacious room intended for research and students' software engineering projects in the context of quadrocopters. The lab supports, e.g., 
trajectory planning with collision avoidance for multiple quadrocopters by providing a domain specific language and an augmented reality (AR) system to visualize important information for the quadrocopters.

\begin{figure}[htb]
    \centering
    \includegraphics[width=0.9\textwidth]{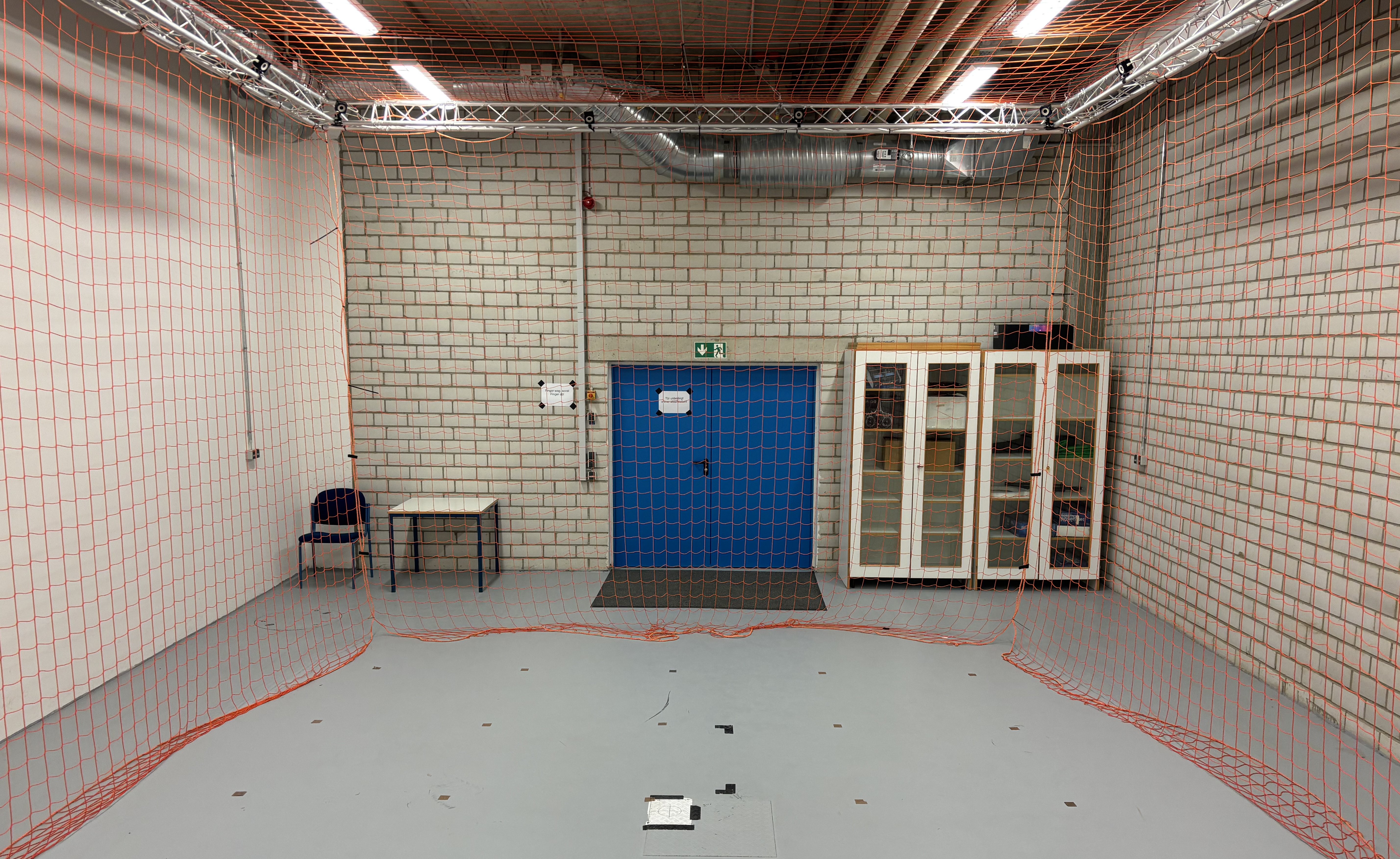}
    \caption{The motion-capturing system in the \textit{Quadcopter Lab}}
    \label{fig:quadcopter_lab}
\end{figure}

The setup includes a flight area (9 x 5 x 4 meters), a motion-capturing system supported by a total of 16 cameras on a traverse near the ceiling
and a Linux machine running all necessary ROS2 infrastructure. The setup is shown in \ref{fig:quadcopter_lab}.

The Robot Operating System 2 (ROS2) is an open-source \textquote{software platform for developing robotics applications}~\cite{Macenski2022RobotOperatingSystem} with widespread applications in both academic and industry settings~\cite{Bonci2023RobotOperatingSystem}.
Despite its name, ROS2 is not a standalone operating system, it is rather an ecosystem or collection of software designed to create modular, distributed and asynchronously communicating units called \textit{nodes}, that usually encapsulate one specific function~\cite{Macenski2022RobotOperatingSystem}.
Running \textit{nodes} can communicate with each other and exchange data asynchronously using different methods. 
Another part of the ROS2 ecosystem is \emph{RViz}~\cite{RVizUserGuide}, a 3D visualizer and \emph{RQt}, a framework for custom user interfaces~\cite{OverviewusageRQt}.

An \textit{OptiTrack}-system \cite{Optitrack} precisely tracks the position and orientation, the so-called \textit{pose},
of any marked objects within an area of approximately 6 x 4 x 3 meters. A marked object refers to an object with special reflective markers attached 
that can be tracked by the cameras using infrared light.
The \textit{OptiTrack}-system utilizes these markers to precisely determine the position of an object in three-dimensional space relative to
a reference "world" frame. The system continuously sends a stamped pose to the Linux machine at a frequency of nearly 100 Hz.
A stamped pose includes the timestamp at which the point was tracked as well as the points' cartesian coordinates and a quaternion for the orientation\footnote{\url{https://docs.ros2.org/galactic/api/geometry_msgs/msg/PoseStamped.html}} of the corresponding object.

The quadrocopters can either be controlled manually by a controller or automatically by so-called mission scripts. The mission scripts are programmed using a simple Lua-based domain-specific language (DSL) presented in \cite{Witte2019HybridEditorFast}. 

The DSL includes several commands like \texttt{takeoff} or \texttt{land}. The most important is \texttt{moveTo(x',y',z',$\psi$)} that directs the quadrocopter to fly from its current position $(x,y,z)$ to the given position $(x',y',z')$, while adjusting its orientation by the angle $\psi$.

For this work, we extended the original language presented by Witte et al.~\cite{Witte2019HybridEditorFast} by convenient functions for creating waypoints for arcs. For example, the function \texttt{arcLeft(n, x, y, z, $\phi$, angle, forward, lateral)} generates $n$ equidistant waypoints from $(x, y, z)$ to the target, which is determined by the left-hand elliptical arc with the \texttt{angle} of an ellipse with the radii \texttt{forward} and \texttt{lateral}. $\phi$ is the angle between the initial direction of the arc and the current orientation of the drone.

%% file: sections/foundatations.tex
\section{Foundations}%
\label{sec:foundatations}

In this section, we briefly formalize terms used throughout this work.
Firstly, we define \emph{paths} and \emph{trajectories} and their properties,
then we formalize their respective planning process and give a definition for \emph{cubic splines} as a way of trajectory interpolation.
Lastly, definitions for different kinds of errors used in the evaluation are given.

We define the \emph{Geometric path} as follows,

\begin{definition}[(Geometric) Path]%
    \label{def:path}
    A path can be defined as a function 
    \begin{align*}
        P: [0, 1] &\rightarrow C  \\
        s &\mapsto p(s)
    \end{align*}
    devoid of any timing information, which satisfies the following constraints~\cite{Pham2015TrajectoryPlanning52}:
    \begin{align*}
        p(0) &= p_{start} \\
        p(1) &= p_{goal}
    \end{align*}
    where $p_{start}$ denotes the first point of the path and $p_{goal}$ the last point of the path.
\end{definition}

where $C$ is called the configuration space, which --- in the context of this work --- is set to $C = \mathbb{R}^3$ as we disregard the orientation of the quadrocopter. Thus, leaving only the spatial three of the usual six degrees of freedom of a quadrocopter.

Then, by adding a time parameterization $s$ that maps the time to a specific section of the path, defined by
\begin{align*}
    s: [t_0, t_{\max}] &\rightarrow [0,1] \\
    t &\mapsto s(t) \\ 
\end{align*}
with $s(t_0) = 0$ and $s(t_{\max}) = 1$, a trajectory is defined as following~\cite{Pham2015TrajectoryPlanning52}:

\begin{definition}[Trajectory]%
    \label{def:trajectory}
    A trajectory T is a path P endowed with a time parameterization s:
    \begin{align*}
        T: [t_0, t_{\max}] &\rightarrow \mathbb{R}^3 \\
        t &\mapsto p(s(t))
    \end{align*}
\end{definition}

More descriptively, \textquote{a trajectory [can be] defined as a sequence of time-stamped locations}~\cite{JoachimGudmundssonComputationalMovementAnalysis}.

In order to retrieve the velocity and the acceleration of the trajectory at each point in time $t$, the trajectory $T$ can simply be derived~\cite{Mellinger2011Minimumsnaptrajectory}.
Then the velocity is defined as $v(t) = \dot{T} = \frac{d}{dt} T$ and the acceleration as $a(t) = \ddot{T} = \frac{d^2}{dt^2} T$.
In addition, we assume for the trajectory planning that the velocity and acceleration at the first and the last point
of the trajectory is set to 0:
\begin{eqnarray*}
    v(t_0) = v(t_{\max}) = 0 \\
    a(t_0) = a(t_{\max}) = 0
\end{eqnarray*}

\begin{definition}[Trajectory smoothness]%
    \label{def:trajectory_smoothness}
    Given a trajectory without excessive or fast-changing accelerations --- meaning the so-called jerk is low, 
    which is defined as the derivate of the acceleration --- this trajectory is called \textbf{smooth}~\cite{Gasparetto2015Pathplanningtrajectory}.
\end{definition}
A low or at least limited \textit{jerk} also has the advantage that the trajectory can be executed faster and with higher accuracy~\cite{Gasparetto2015Pathplanningtrajectory}.

Based on the definitions of a path and a trajectory, the terms \textit{path planning} and \textit{trajectory planning}
are defined as:

\begin{definition}[Path planning]%
    \label{def:path_planning}
    Path planning is the process of generating a geometric path (see Definition~\ref{def:path}) from a start to a goal point,
    passing through pre-defined waypoints~\cite{Gasparetto2015Pathplanningtrajectory}.
\end{definition}
\begin{definition}[Trajectory planning]%
    \label{def:trajectory_planning}
    Trajectory planning is the process of endowing a geometric path (see Definition~\ref{def:path}) with time information~\cite{Gasparetto2015Pathplanningtrajectory}.
\end{definition}

As the geometric path is already defined in the demonstration, further consideration of path planning is unnecessary.
In practice, different approaches for trajectory planning are employed focusing on speed, efficiency, or minimal jerk~\cite{Gasparetto2015Pathplanningtrajectory}, however, in this work, the focus is on generating a smooth and precise trajectory.
Therefore, the (natural) cubic spline is presented as this is a simple, smooth, and accurate trajectory planning method. Cubic splines are used to create smooth curves fitting through several waypoints and are defined as follows:

\begin{definition}[Cubic spline]%
    \label{def:cubic_spline}
    Given a set $X = \{(x_1, y_1), \dots, (x_n, y_n)\}$ of point-pairs,
    a cubic spline is a function $S: [x_1, x_n] \mapsto \mathbb{R}$ defined piece-wise by cubic functions~\cite{McKinley1998Cubicsplineinterpolation}:
    \begin{equation*}
        S(x) = \begin{cases}
            s_1(x) ~~~~~~~~~~ x_1 \leq x \leq x_2 \\
            \dots \\
            s_{n-1}(x) ~~~~~~ x_{n-1} \leq x \leq x_n \\
        \end{cases}
    \end{equation*}
    where $s_i$ is a polynomial of third degree given by:
    \begin{equation*}
        s_i(x) = a_i{(x - x_i)}^3 + b_i{(x - x_i)}^2 + c_i(x - x_i) + d_i
    \end{equation*}
    A cubic spline $S$ is a natural cubic spline, if $s_1''(x_1) = s_{n-1}''(x_n) = 0$. %
\end{definition}

From this, we can derive the following properties for a cubic spline function~\cite{McKinley1998Cubicsplineinterpolation}:
\begin{enumerate}
    \item $\forall x_i \in X: S(x_i) = y_i$, i. e. $S$ interpolates all (way)points $y_i$ exactly
    \item $S \in C^2$, i. e. $S$ is twice continuously differentiable meaning $S(x)$, $S'(x)$, $S''(x)$ are continuous on [$x_1$, $x_n$] 
\end{enumerate}

Using cubic splines to interpolate trajectories can lead to errors with regard to the demonstration --- the so-called \emph{interpolation error}.
A typical method to determine this error is to calculate the position error vectors~\cite{Hoppe2019Dronosflexibleopen,Luis2016Designtrajectorytracking,JoachimGudmundssonComputationalMovementAnalysis}.
The position error vector for a specific point in time $t$ is defined as 
\begin{eqnarray}
    e_{pos}(t) &=& P(t) - T(t) \\
    &=& {[P_x(t) - T_x(t), P_y(t) - T_y(t), P_z(t) - T_z(t)]}^T
\end{eqnarray}
where $P(t)$ describes the point of the demonstrated flight path at time $t$ and $T(t)$ the corresponding point of the trajectory at the same time $t$.

In addition, there may be an error between the theoretically planned trajectory and the actually flown trajectory in the real world, potentially caused by external influences, such as a ventilation, but as it is not directly related to the trajectory planning process, this type of error is not considered further in this work.
We also limit the calculation of errors to the planned trajectory as opposed to the edited trajectory, as the latter would not necessarily represent an error of the trajectory planning process itself.

These previously mentioned position error vectors are the foundation of further error analysis, wherefore the root-square-mean-error (RSME)~\cite{Xu2019Robottrajectorytracking,Foehn2021Timeoptimalplanning,Hoppe2019Dronosflexibleopen}
and the mean-absolute-error (MAE) are commonly used. The advantage of the MAE over RSME is that it emphasizes outliers less and can be interpreted directly.
The RSME and MAE can be calculated as follows\footnote{In both~\ref{eq:rsme} and~\ref{eq:mae}, $||\bullet||_2$ denotes the Euclidean norm.}~\cite{Chai2014Rootmeansquare}:
\begin{equation}
    \label{eq:rsme}
    RSME = \sqrt{\frac{1}{n} \sum_{i=1}^{n} ||e_{pos}(t_i)||_2^2}
\end{equation}
\begin{equation}
    \label{eq:mae}
    MAE = \frac{1}{n} \sum_{i=1}^{n} ||e_{pos}(t_i)||_2
\end{equation}
where $n$ is the number of recorded points making up the demonstrated flight path and $t_i$ is the time of the $i$-th recorded point.

However, a non-zero interpolation error is not necessarily problematic: Even though during trajectory planning physical limitations are taken into account and jitter in demonstrations is ignored as much as possible, the errors are still calculated between the raw demonstration and planned trajectory.

%% file: sections/drawjectory_workflow.tex
\section{Concept and Implementation of the \textit{Drawjectory} workflow}
\label{sec:drawjectory_workflow}

To solve the problem of a precise and intuitive manual trajectory planning for quadrocopters,
we propose the \textit{Drawjectory} workflow.
This workflow comprises four distinct phases, depicted in \ref{fig:workflow}, that build upon each other:%
\vspace*{-1.5em}
\begin{wrapfigure}[6]{r}{0.5\textwidth}%
	\begin{center}
		\includegraphics[width=0.47\textwidth]{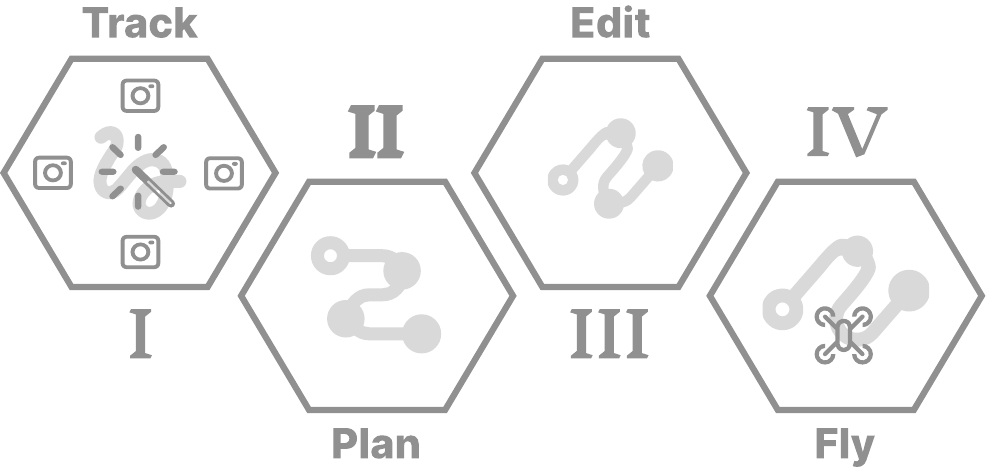}
		\captionof{figure}{The \textit{Drawjectory} phases}
		\label{fig:workflow}
	\end{center}
\end{wrapfigure}%
\begin{enumerate}
    \item Track the desired flight path of the quadrocopter using a gesture wand.
    \item Plan a smooth, feasible trajectory based on the demonstration.
    \item Edit the planned trajectory, by shifting, rotating or linearly scaling it, or by moving certain points.
    \item Process trajectory to control the quadrocopter.
\end{enumerate}

The objective of this workflow is to demonstrate the desired trajectory intuitively in the first phase.
The adapted trajectory planning in phase two ignores the noise from the user's demonstration by creating a smooth, feasible trajectory.
To address the issue of the limited space in the quadrocopter laboratory, or of slightly deviating demonstrations, we introduce the third phase.
This phase provides the user with the option to edit the trajectory, by either moving certain points or shift, scale or rotate the trajectory relative to the first
point of the trajectory\footnote{In the \textit{Quadcopter Lab}
it is not possible to scale the trajectory arbitrarily, as the quadrocopters need a connection to the \textit{OptiTrack}-system and Linux machine.}.

As motivation for this phase, our programming by demonstration approach is restricted to the space covered by the tracking system. 
Consequently, it is necessary to enlarge the planned route at a later stage and, if necessary, to rotate or move it relative to a calibration point.
If the user later realizes that parts of the flight were unintentional or too inaccurate, moving the waypoints is useful.

In order to test the \textit{Drawjectory} workflow, we created a RQt plugin called \texttt{Drawjectory Control Panel} that serves as a proof-of-concept implementation.
More details on the architecture, the implementation and integration of the \texttt{Drawjectory Control Panel} into the \textit{Quadcopter Lab} can be found in \cite{Alkewitz2024HowDrawjectoryTrajectory}.

\subsection{Tracking of the Gesture Wand}
\label{ssec:tracking_gesture_wand}

The first phase of the workflow is the \textit{demonstration of the flight path}, consisting of the following steps:
i) Tracking an object that determines the flight path,
and ii) the saving, editing and visualizing of the demonstrated flight path.

\begin{wrapfigure}[15]{r}{0.35\textwidth}
    \centering
    \includegraphics[width=0.35\textwidth]{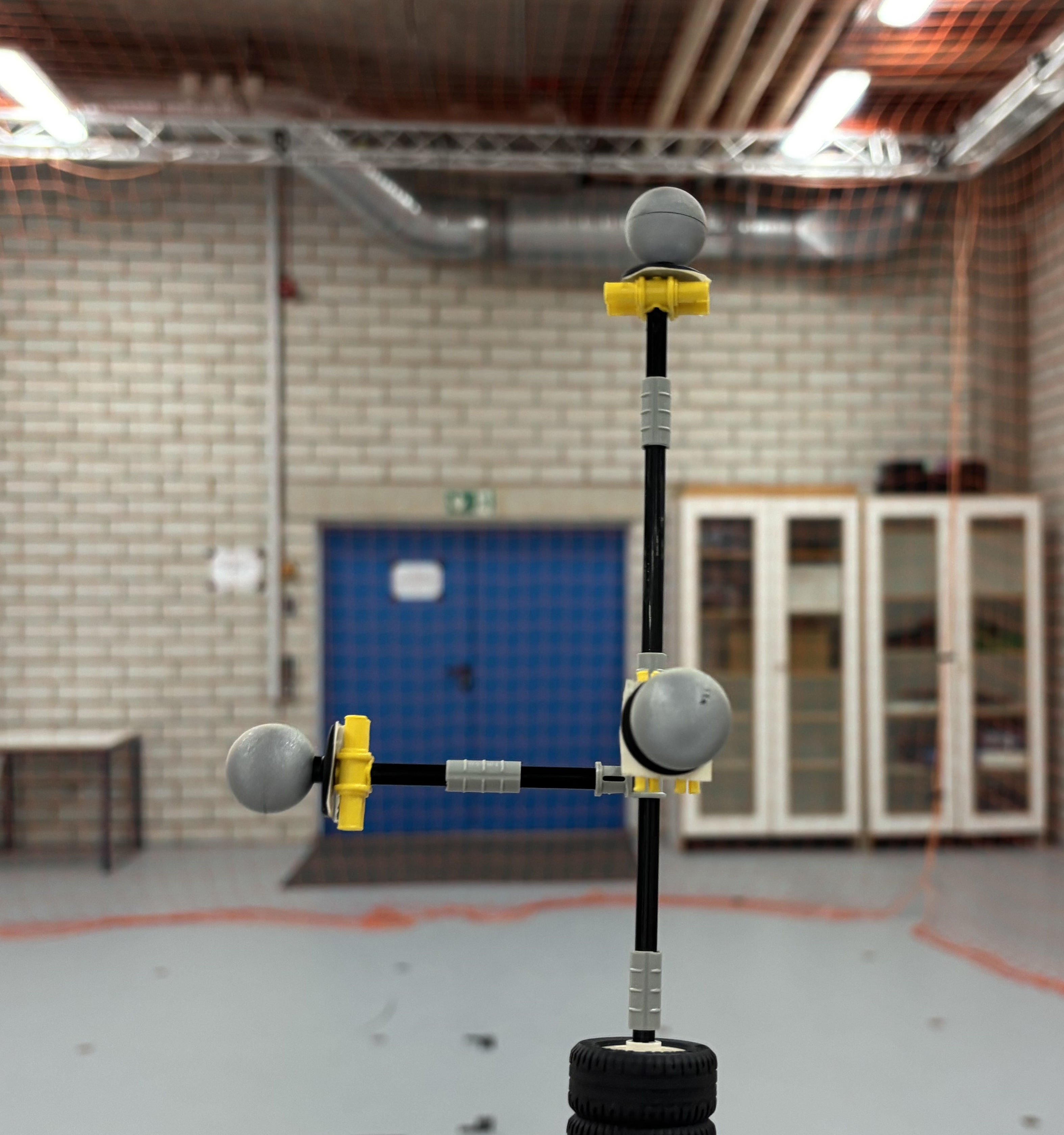}
    \caption{The gesture wand with grey reflective markers.}
    \label{fig:gesture_wand}
\end{wrapfigure}
To demonstrate the flight path of the quadrocopter, we use a custom gesture wand.
As shown in \ref{fig:gesture_wand}, the wand is a stick with three reflective markers attached to it,
aligned with the axes of a Cartesian coordinate system, that is tracked by the \textit{OptiTrack}-system.
It should be noted that the demonstration may be imprecise, or not determined at all,
as the tracking cameras are mounted on a truss near the ceiling and can only observe a limited area within the quadrocopter laboratory.
However, this area is marked on the floor and working solely in this area reduces the likelihood of poor quality tracking (see \ref{fig:quadcopter_lab}).

The position of the gesture wand is sent to the \texttt{Drawjectory Control Panel} only approximately every $10~ms$, 
although this is not a serious issue as points are selected to interpolate a trajectory later.
These recorded points make up the demonstrated flight and are the basis for later trajectory planning.

\begin{figure}[htb]
    \centering
    \includegraphics[width=0.8\textwidth]{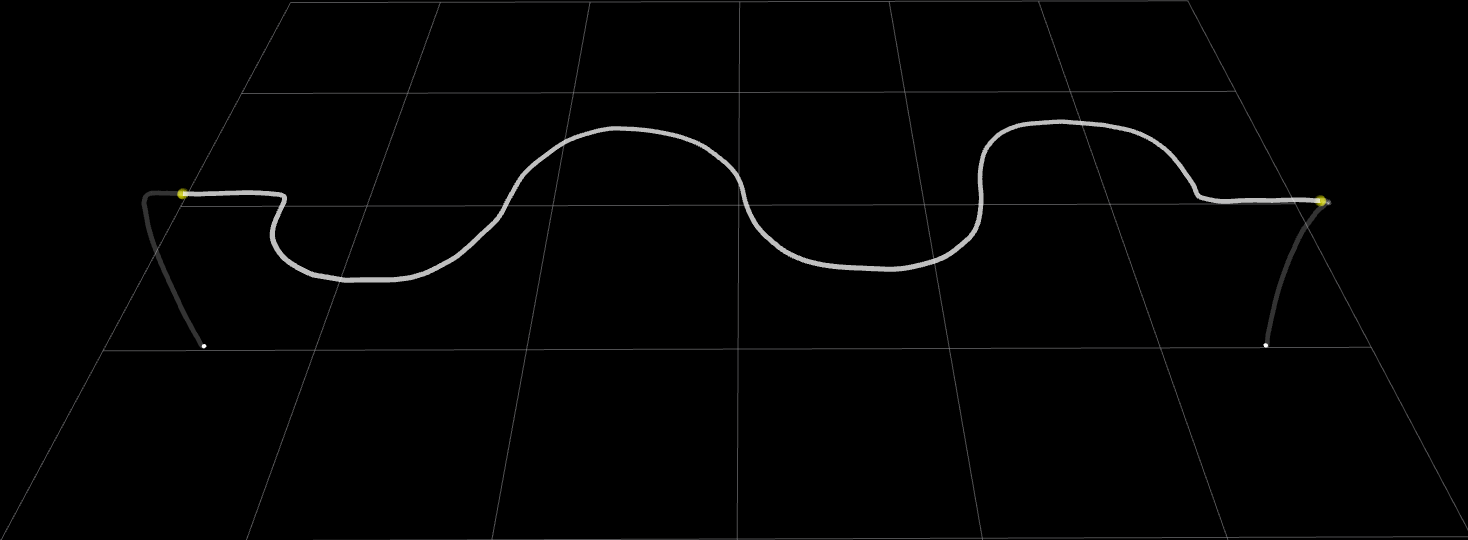}
    \caption{Visualization of the demonstrated, trimmed flight path in \textit{RViz}, whereas the two yellow spheres mark the start and end point of the trimmed flight path.}
    \label{fig:demonstrated_flight_path_visualization}
\end{figure}

In addition, the \textit{Drawjectory} workflow includes the option to trim and visualize the recorded flight path, exemplified in \ref{fig:demonstrated_flight_path_visualization}.
\textquote{Trimming} the flight path is the process of redefining the actual start and end points of the flight path so that
all points before the start point and all points after the endpoint are ignored when planning the trajectory.
One advantage of trimming is that the user is allowed to cut out parts at the beginning or end of the demonstration that contained errors rather than restarting the tracking,
or cut out a section of the demonstrated flight path that is only necessary, for instance, if the flight path is only to extend an automatically planned trajectory.

It is advisable to allow the user of the \textit{Drawjectory} workflow to save these recorded points
in order to reload them later or elsewhere if necessary, e. g. to proceed the workflow in a different environment. 
Therefore, the \texttt{Drawjectory Control Panel} provides methods
to save and (re-)load a list of $n$ recorded points in an unambiguous format.

\subsection{Adapted Trajectory Planning}
\label{ssec:adapted_trajectory_planning}

Based on the truncated demonstrated flight the next phase of the \textit{Drawjectory} workflow is \textit{to plan or reconstruct the trajectory}.
This is typically achieved by sampling points from the demonstration and interpolating
a path passing through the sampled points, with the aim of including the omitted points as well~\cite{Aleotti2005Trajectoryreconstructionnurbs,Hwang2003Mobilerobotsyour,Melchior2012Graphbasedtrajectory,Zimmermann2023Twosteponline}.
The reason for the sampling is to reduce the noise of the demonstration and create smoother curves by interpolation.

\subsubsection{Selection of the Waypoints}
\label{sssec:waypoint_selection}

A number of sampling strategies exist to extract waypoints from a sequence of recorded points. 
These include random sampling, equidistant sampling and corner detection \cite{Hwang2003Mobilerobotsyour}, 
whereas random and equidistant sampling are techniques developed in-house.
In the proof-of-concept implementation of the workflow, we employed random and equidistant sampling,
as random sampling is a simple and fast-to-implement approach, while equidistant sampling regarding the spatial distance is
commonly used for obtaining waypoints for the interpolation \cite{BochkanovSplineinterpolationfitting}. 
The principal disadvantage of these methods is the lack of focus on \textquote{significant} points, such as points in sharp curves, 
which is why the corner detection algorithm exists.
Nevertheless, this algorithm is too complex for a proof-of-concept implementation and reduces noise to little \cite{Hwang2003Mobilerobotsyour}.

\begin{wrapfigure}[13]{L}{0.33\textwidth}
	\vspace*{-4em}
    \begin{minipage}{0.32\textwidth}
        \begin{algorithm}[H]
            \caption{Equidistant waypoint selection}\label{alg:equidistant_sampling}
            \begin{algorithmic}
            \Require $n \geq 2$
            \State $W \gets \{\}$
            \State $d \gets round(\frac{|P| - n}{n - 1})$
            \State $i \gets 0$
            \While{$i < |P| - 1$}
                \State $W \cup p_i$
                \State $i \gets i + d + 1$   
            \EndWhile
            \State $W \gets \{p_l\}$
            \end{algorithmic}
        \end{algorithm}                
    \end{minipage}
\end{wrapfigure}
In general, the strategies used in the \textit{Drawjectory} workflow are required to include i) the first and last point $P_0$ and $P_l$
of the trimmed demonstrated flight path $P$ and ii) a specified number of waypoints $n$.
Random sampling retrieves a point $p_i$ from the sequence of recorded points $P$ (except the already chosen points) as long as the number of desired waypoints is not reached. 
An alternative approach is to select the waypoints in an equidistant manner. Algorithm \ref{alg:equidistant_sampling} introduces a technique of 
selecting the waypoints in such a way that there are approximately the same number of omitted points between each adjacent waypoint.
Assuming that the time interval between each recorded point $p_i$ is perfectly equal, this technique is equivalent
to selecting the waypoints regarding a certain time interval, so use equidistant sampling in terms of time.
Consequently, the \textquote{distance}, i.e. the number of omitted points between two waypoints and not the spatial distance, must be determined by the following formula:
\begin{equation*}
    d = \frac{\text{number of points in sequence} - \text{number of waypoints}}{\text{number of waypoints} - 1}
\end{equation*}
It describes the points to be omitted ($|P| - n$) are split into ($n - 1$) sections, as there are this many sections between the waypoints.
The rationale behind this sampling technique is based on the assumption that a user demonstrates more complex sections, such as a sharp curve, of the trajectory at a slower pace and simpler ones at a faster pace. 
Consequently, there is a higher density of points in sections that are more complex (to interpolate), which in turn leads to a greater number of waypoints in these sections, which are used for interpolation.
This sampling technique therefore produces results that are comparable to those obtained through corner detection, but is more likely to reduce noise, provided that not too many waypoints are selected.

\subsubsection{Interpolation of the Trajectory}
\label{sssec:interpolation}

For interpolating a trajectory based on given waypoints, there exist a couple of methods, such as 
Bezier curves or NURBS, as presented in \ref{sec:related_work}.
The most basic idea, however, is to linearly interpolate, i.e. connect the waypoints with straight lines. 
The method's principal advantage is its straightforward implementation and the fact that it creates exact lines.
However, this method also has significant drawbacks, including its inefficiency as it creates infinite curvature at the waypoints, which necessitates the 
quadrocopter's halt at the waypoints to precisely adhere to the trajectory \cite{Mellinger2011Minimumsnaptrajectory}.

Another commonly used approach to planning the trajectory are polynomial piecewise functions, such as natural cubic splines, which were
introduced in \ref{sec:foundatations}. Cubic splines offer the advantage of a simple and expedient calculation, while their smoothness remains quite high
and they pass exactly through the waypoints in contrast to B-splines \cite{Usenko2017Realtimetrajectory}.
As cubic splines are only capable of interpolating two-dimensional functions and a trajectory requires the interpolation of a four-dimensional function,
the trajectory cannot be interpolated directly but need to be decoupled. 
This entails interpolating the trajectory not as a whole, but separately for the $x$, $y$ and $z$ coordinates, and subsequently coupling these
coordinates \cite{Hehn2011Quadrocoptertrajectorygeneration}.

Given the sequence of waypoints, denoted as $W := p_0, ..., p_n$, where each point $p_i$ was recorded at time $t_i$,
three sets of time-coordinate pairs, one for each of the spatial axes $x$, $y$ and $z$, are defined: $X := \{(t_0, p_{0,x}), ..., (t_n, p_{n,x})\}$, $Y := \{(t_0, p_{0,y}), ..., (t_n, p_{n,y})\}$ and $Z := \{(t_0, p_{0,z}), ..., (t_n, p_{n,z})\}$. 
These time-coordinate pairs determine the cubic splines $S_x$, $S_y$ that in turn define the trajectory $T$ as follows: 
\begin{equation*}
    T(t) = [S_x(t), S_y(t), S_z(t)]^T
\end{equation*} 
To fly the trajectory, it is necessary to obtain control points from the trajectory, as well as the
velocity and acceleration at certain points. Therefore, the splines and their derivatives are evaluated at timestamps $t \in [t_0, t_{max}]$.

\subsubsection{Feasibility Constraints for Trajectories}
\label{sssec:trajectory_feasibility}

As either the demonstration or the interpolation can result in a trajectory that
is not flyable, also called \textit{not feasible}, it is necessary to fulfill certain constraints before the trajectory is flown.

The first constraint is the size of the \textit{Quadcopter Lab}, or more precisely the area in which 
the quadrocopters are allowed to fly.
As previously stated in \ref{sec:quadcopter_lab}, the area in question has a size of 6 x 4 x 3 meters.
Consequently, each point $p_i$ of the trajectory $T$ must be within this area, i.e. 
\begin{equation*}
\forall p_i \in T: 0 \leq p_{i,x} \leq 6 \land 0 \leq p_{i,y} \leq 4 \land 0 \leq p_{i,z} \leq 3
\end{equation*}
otherwise the demonstration must be repeated.

The controllers of the quadrocopter impose the second constraint, as they limit the
velocity and acceleration of the quadrocopter \cite{Mueller2015computationallyefficientmotion}.
In general, the velocity $v(t)$ and acceleration $a(t)$ must fulfill the following constraints:
\begin{eqnarray*}
    ||v|| = \sqrt{v_x^2 + v_y^2 + v_z^2} \leq v_{max} \\
    ||a|| = \sqrt{a_x^2 + a_y^2 + a_z^2} \leq a_{max} 
\end{eqnarray*}

In the context of the \textit{Quadcopter Lab}, experimenting with different limits for the velocity and 
acceleration revealed that $v_{max} = 1.5~\frac{m}{s}$ is appropriate. Additionally, it is sufficiently low
to prevent the quadrocopter from reaching a critical acceleration $a_{max}$, which would result in a deviating flown trajectory. 

\subsubsection{Visualization of Trajectory}
\label{sssec:trajectory_visualization}

\begin{figure}[b]
    \centering
    \includegraphics[width=0.8\textwidth]{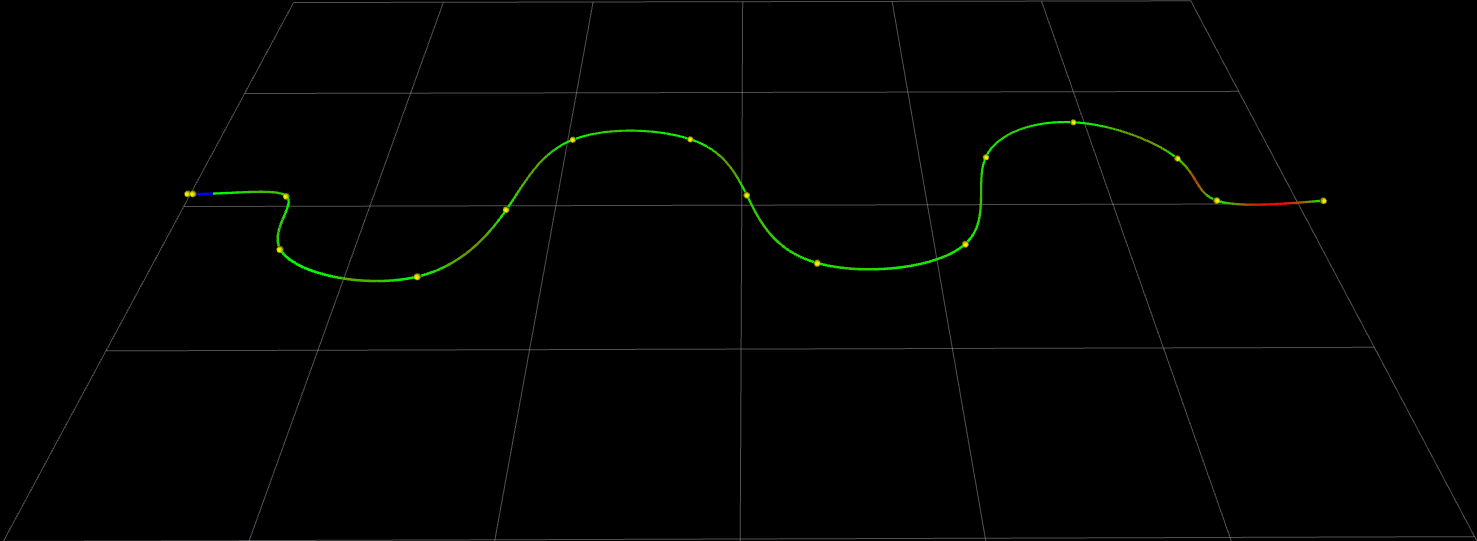}
    \caption{Visualization of a trajectory, colored according to the interpolation error}
    \label{fig:visualization_trajectory_with_error}
\end{figure}

In the case of the proof-of-concept implementation, a line visualizes the computed trajectory in a manner analogous to the demonstrated flight path. 
This visualization is extended by yellow spheres representing the waypoints. The trajectory is colored with a color gradient, 
ranging from green (minor deviation) to red (major deviation), indicating the interpolation error (see \ref{fig:visualization_trajectory_with_error}).

\subsubsection{Editing Planned Trajectory}
\label{sssec:edit_trajectory}

Once the trajectory planning phase is complete, the user may \textit{edit the trajectory}.
There are several operations, which may or may not be applied and which may be combined, and which 
therefore allow a trajectory to be edited arbitrarily, as long as these operations do not produce a trajectory control point that violates
the feasibility constraints (cf. \ref{sssec:trajectory_feasibility}).

One operation is to manipulate the waypoints of the trajectory within a three-dimensional virtual environment, such as \textit{RViz}.
Additionally, the trajectory may be shifted by an offset, rotated by an angle around the first point in the xy-plane
or linearly scaled by a factor relative to the first trajectory point\footnote{See \cite{Deakin19983Dcoordinate} for more information about how these
operations can be defined.}. 

\begin{figure}[tb]
	\centering
	\includegraphics[width=0.8\textwidth]{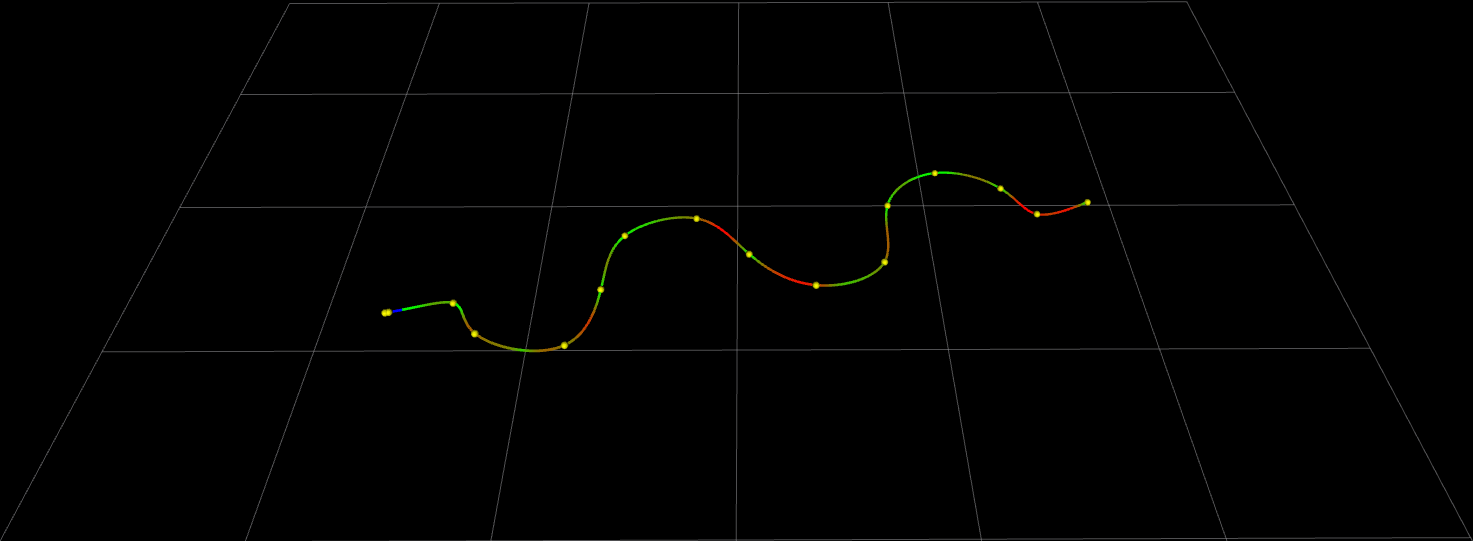}
	\caption{Visualization of an shifted, scaled, rotated trajectory with moved waypoints.}
	\label{fig:visualization_edited_trajectory}
\end{figure}
\ref{fig:visualization_edited_trajectory} illustrates an example of an edited trajectory.
As can be seen, several parts of the trajectory are red, indicating that these section deviate the most
from the demonstration. However, as the lines are now relatively close to half circles, these adaptions probably 
match the intended flight path more closely.

%% file: sections/evaluation.tex
\section{Evaluation}
\label{sec:evaluation}

To evaluate the \textit{Drawjectory} workflow, we applied the workflow to a couple of scenarios and measured
the planning time as well as the accuracy. This experiment should provide some insight into whether the \textit{Drawjectory} workflow 
is faster - i.e. easier to use - and at what cost, namely loss of accuracy compared to programming the trajectory, this possible acceleration comes.
We provide a package online with the material
used for the evaluation, namely the raw programming, demonstration and trajectory data,
the evaluation scripts and detailed scenario descriptions~\cite{alkewitz_2024_13992727}.

\paragraph{Unit Testing and Test Coverage}
As the correctness of the trajectory planning process is important too, this evaluation is complemented by unit tests.
Therefore, 45 test cases with multiple conditions have been written that test the implementation of the \textit{Drawjectory} workflow automatically.
These test cases achieve a statement coverage of 60.2\% and excluding the parts responsible for the GUI, it increases to 95.5\%.

\subsection{Experiment Setup}
\label{ssec:experiment_setup}

In general, the experiment (setup) consists of three phases: i) the definition and specification of the scenarios, 
ii) conducting the experiment by realizing the scenarios using two input modalities and measuring the planning time
as well as the accuracy, iii) evaluation and analysis of the measurements. 
The experiment was conducted by one author, an experienced user of both input modalities. While programming and demonstrating the trajectories, 
the author was allowed to use the scenario specification to read the instructions. Corrupted demonstrations, 
such as demonstrations with missing tracking data or trajectories that did not conform to the specification, have been repeated.

\paragraph{Scenarios}

We designed a total of 16 scenarios, divided into three categories of increasing complexity (cf. \ref{tab:similarities}):
\begin{itemize}
	\item \textit{Planar geometric figures} (5), such as a square or an ellipse
	\item \textit{Planar non-trivial figures} (6), such as simulating the inspection of rooms
	\item \textit{3D figures} (5), such as a spiral or staircase
\end{itemize}
In general, the scenarios were designed to represent complete (second and third group) or at least components (first group) of possible
real-world scenarios. However, the scenarios were intended to be solvable in time, meaning they should not be too complicated.
The different scenarios should test, how well trajectories containing sharp turns, smooth curves, long lines or alternating heights
can be planned.

\begin{wrapfigure}[6]{R}{0.4\textwidth}
	\vspace*{-3.5em}
	\tdplotsetmaincoords{60}{175}
	\begin{tikzpicture}[tdplot_main_coords, scale = 0.75] 
		\draw[->] (0,0,0) -- (6,0,0) node[anchor=north west]{x};
		\draw[->] (0,0,0) -- (0,2,0) node[anchor=north east]{y};
		\draw[->] (0,0,0) -- (0,0,1) node[anchor=south]{z};
		
		\begin{scope}[canvas is xy plane at z=1.0]
			\draw[-, line width=0.5mm, rounded corners=2pt] (0.5,2,0) -- (1,2,0);
			\draw[line width=0.5mm] (1, 2, 0) arc (180:360:0.5);
			\draw[line width=0.5mm] (3, 2, 0) arc (0:180:0.5);
			\draw[line width=0.5mm] (3, 2, 0) arc (180:360:0.5);
			\draw[line width=0.5mm] (5, 2, 0) arc (0:180:0.5);
			\draw[-, line width=0.5mm, rounded corners=2pt] (5,2,0) -- (5.5,2,0);
		\end{scope}
		\cuboid{1.5}{2}{1}{0.125}
		\cuboid{2.5}{2}{1}{0.125}
		\cuboid{3.5}{2}{1}{0.125}
		\cuboid{4.5}{2}{1}{0.125}
	\end{tikzpicture}
	\vspace*{-1ex}
	\captionof{figure}{Visualization of the scenario \textit{Virtual slalom}, whose realization was shown in \ref{sec:drawjectory_workflow}.}
	\label{fig:slalom}
\end{wrapfigure}

Each scenario is precisely specified and described, complemented by a visualization to illustrate the description, as exemplary shown in \ref{fig:slalom}.

\paragraph{Input Modalities (IV)}

The scenarios were realized using two different input modalities (independent variable):
\begin{itemize}
    \item \textit{\textbf{Programming}} using an extended version of the Domain-specific language (DSL) of the \textit{Hybrid Editor} \cite{Witte2019HybridEditorFast}, already mentioned in the \ref{sec:quadcopter_lab}.
    \item \textit{\textbf{Demonstration}} of the trajectory in the real-world, in this case in the \textit{Quadcopter Lab} with the \textit{OptiTrack} system.
\end{itemize}

Since the \textit{Hybrid Editor} does not directly support the planning and visualization of trajectories, the programmed flight path was sent to the \texttt{Drawjectory Control Panel}
as substitute of the tracking data. Hence, the further planning process, including trimming the flight path, determining the number of waypoints and starting the
trajectory planning is identical, only the input method is different. The resulting trajectory is displayed in \textit{RViz} 
to evaluate whether the demonstration or programming was correct, complete and meets the specification.

\paragraph{Measurements (DV)}

To indirectly evaluate the usability of the \textit{Drawjectory}, we measured the time required to create a trajectory according to the scenario description, as the required time is an indicator for the usability.
The planning time was measured automatically by pressing two buttons on the \texttt{Drawjectory Control Panel}.
Since the planning time does not include the time required to actually plan the trajectory, the time is not dependent on the system on which it is determined.

In order to asses the accuracy of the demonstrated trajectories, we also measured the interpolation error as well as a collection of widely used trajectory similarity measurements.
As the demonstrated and the programmed trajectory are not necessarily the same length or duration, we cannot simply calculate the RSME or MAE of the demonstrated trajectory compared to the programmed trajectory.
That is why we analyzed the demonstrated trajectories using measurements that can determine the similarity of two different long trajectories.
As a comparison value for the trajectory calculated on the basis of the demonstration, we use the programmed flight path, which represents the intended flight path.

The \textit{Hausdorff} distance measures the maximum point-to-trajectory distance, similar to the \textit{Discrete Frechet} distance that is defined as
\textquote{the maximum distance between any matched point pairs}. The \textit{Dynamic Time Warping} (DTW) algorithm finds an optimal matching between the points
of two trajectories by minimizing the sum of point-to-point distances, which is called DTW distance \cite{Toohey2015Trajectorysimilaritymeasures}. Since the DTW distance
is depend on the length of the two trajectories, we also consider a normalized DTW distance, which is determined by dividing the DTW distance
by the maximum length of two trajectories. This measure overestimates the average distance between two matched trajectory points \cite{Chang2024Trajectorysimilaritymeasurement}.
In general, the smaller the distances, the more similar the trajectories are.

\subsection{Results}
\label{ssec:results}

\paragraph{Interpolation Error}

\begin{figure}[htb]
    \begin{minipage}{0.48\textwidth}
        \centering
        \includegraphics[width=0.85\textwidth]{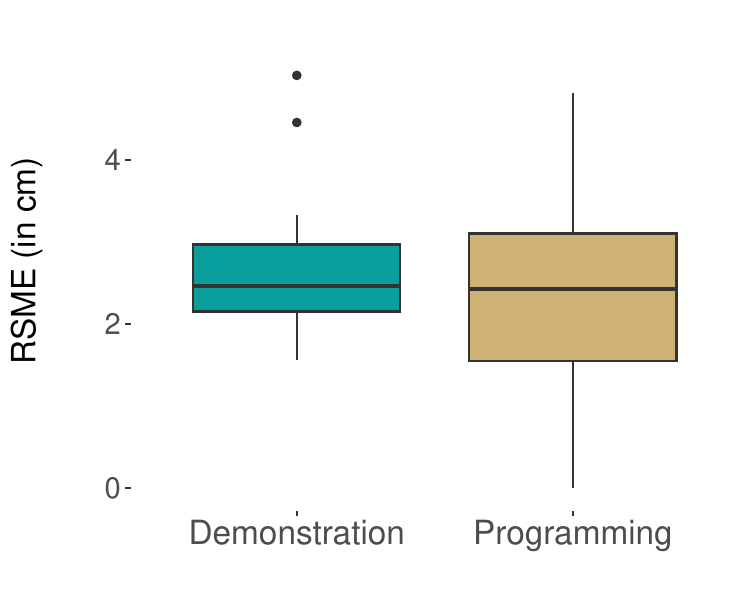}
    \end{minipage}
    \hfill
    \begin{minipage}{0.48\textwidth}
        \centering
        \includegraphics[width=0.85\textwidth]{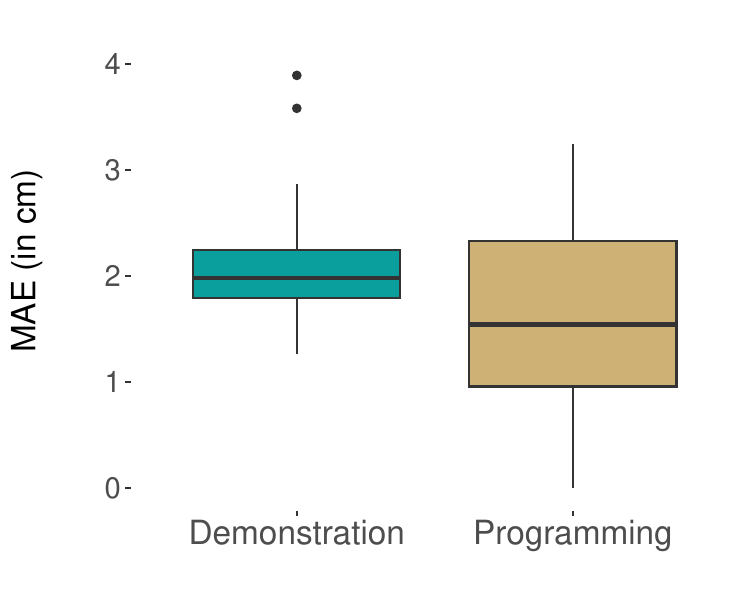}
    \end{minipage}
    \caption{Interpolation errors RSME (left) and MAE (right) for the two input modalities \textit{Demonstration} and \textit{Programming}.}
    \label{fig:interpolation_errors}
\end{figure}

Measuring the interpolation error from both inputs, the demonstration and the programmed flight path, revealed
that the interpolated trajectory is quite similar to the input, as can been see in \ref{fig:interpolation_errors}.

The average RSME of trajectories planned from demonstrations is 0.027 (SD = 0.0093), in comparison to 0.0232 (SD = 0.014) for programmed flights.
Regarding the MAE, the results are very similar: 0.0217 (SD = 0.0073) for the demonstration and 0.0158 (SD = 0.0098) for the programming.
This can be interpreted as an average deviation of 2.17 cm respectively 1.58 cm from the trajectories to their input flight path.
These results show that demonstrating the trajectories also generates sections that cannot be flown directly or that contain noise. 
However, it can be seen that this is of the same order of magnitude as programming and that the two input modalities are therefore comparable. 

\paragraph{Planning Time}

Averaged over all scenarios, using the demonstration approach requires a significantly lower planning time of 75.2 seconds (SD = 18.39) compared to
programming the trajectories with a planning time of 153.86 seconds (SD = 96.96).

\begin{figure}[tb]
	\centering
	\includegraphics[width=0.9\textwidth]{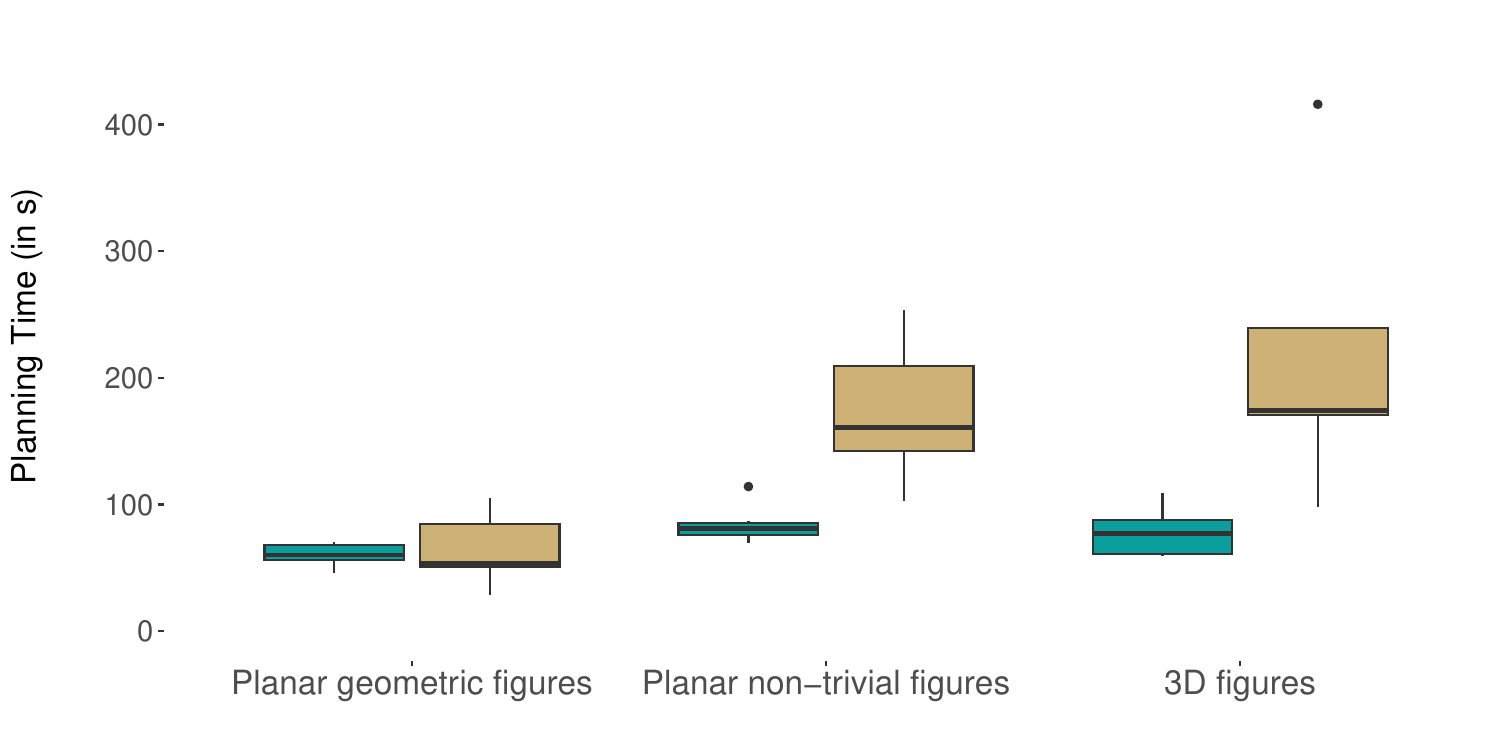}
	\caption{The trajectory planning times for the three scenario groups.}
	\label{fig:planning_time}
\end{figure}

As can be seen in \ref{fig:planning_time}, the required planning time for demonstrating a scenario does not show a strong increase for more complex scenarios
(from 60.2 seconds for geometric figures to 79.0 seconds for complex 3D figures on average). 

On the other hand, the time required to program a trajectory increases as the scenarios become more complex, 
especially if the trajectory is not planar (from 64.7 seconds for geometric figures to 219.7 seconds for complex 3D figures on average).
This indicates that, probably due to its greater intuitiveness, the \textit{Drawjectory} workflow saves time, in particular, in more complex scenarios. 

\paragraph{Trajectory Similarity}

\begin{figure}[htb]
	\begin{minipage}{0.48\textwidth}
		\includegraphics[width=0.95\textwidth]{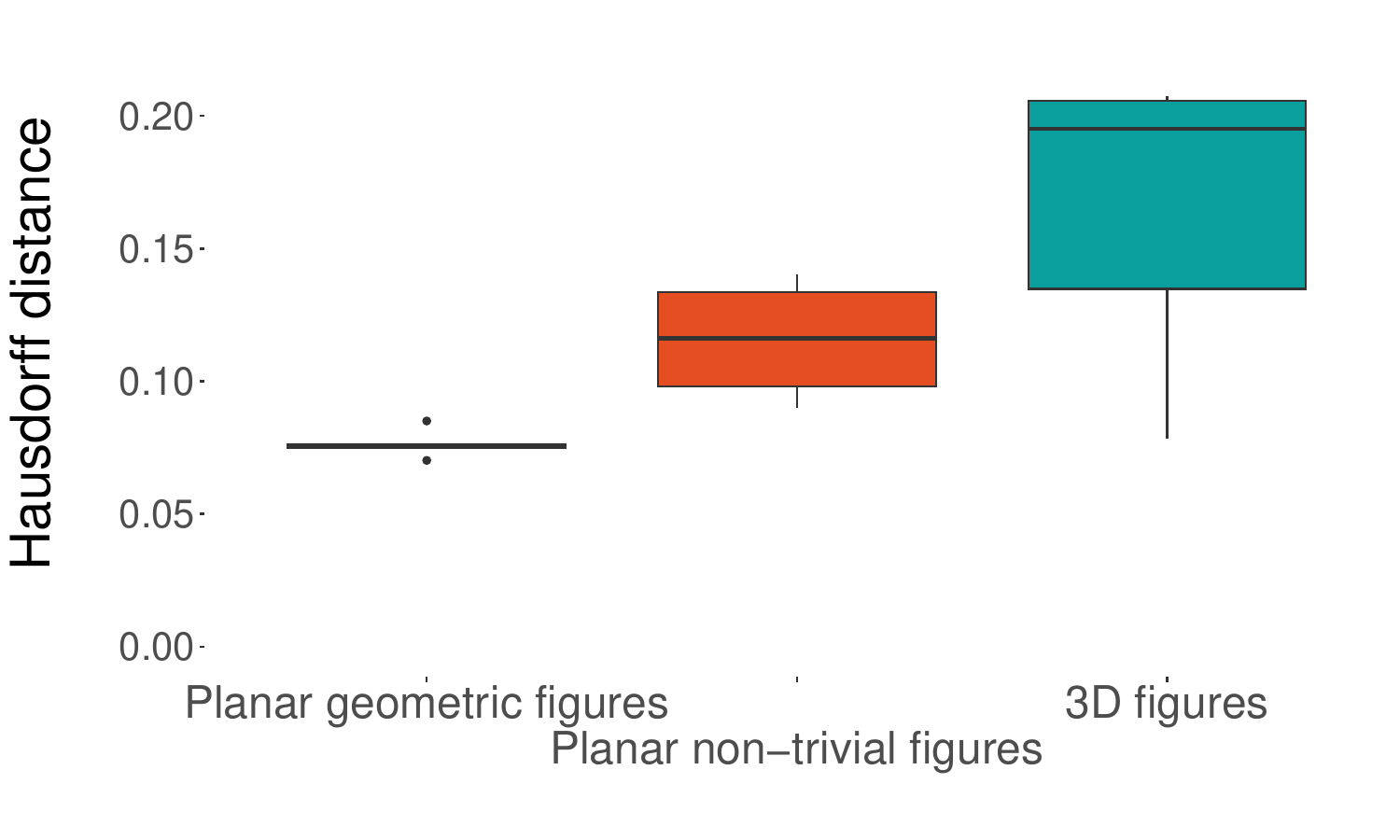}
		\includegraphics[width=0.95\textwidth]{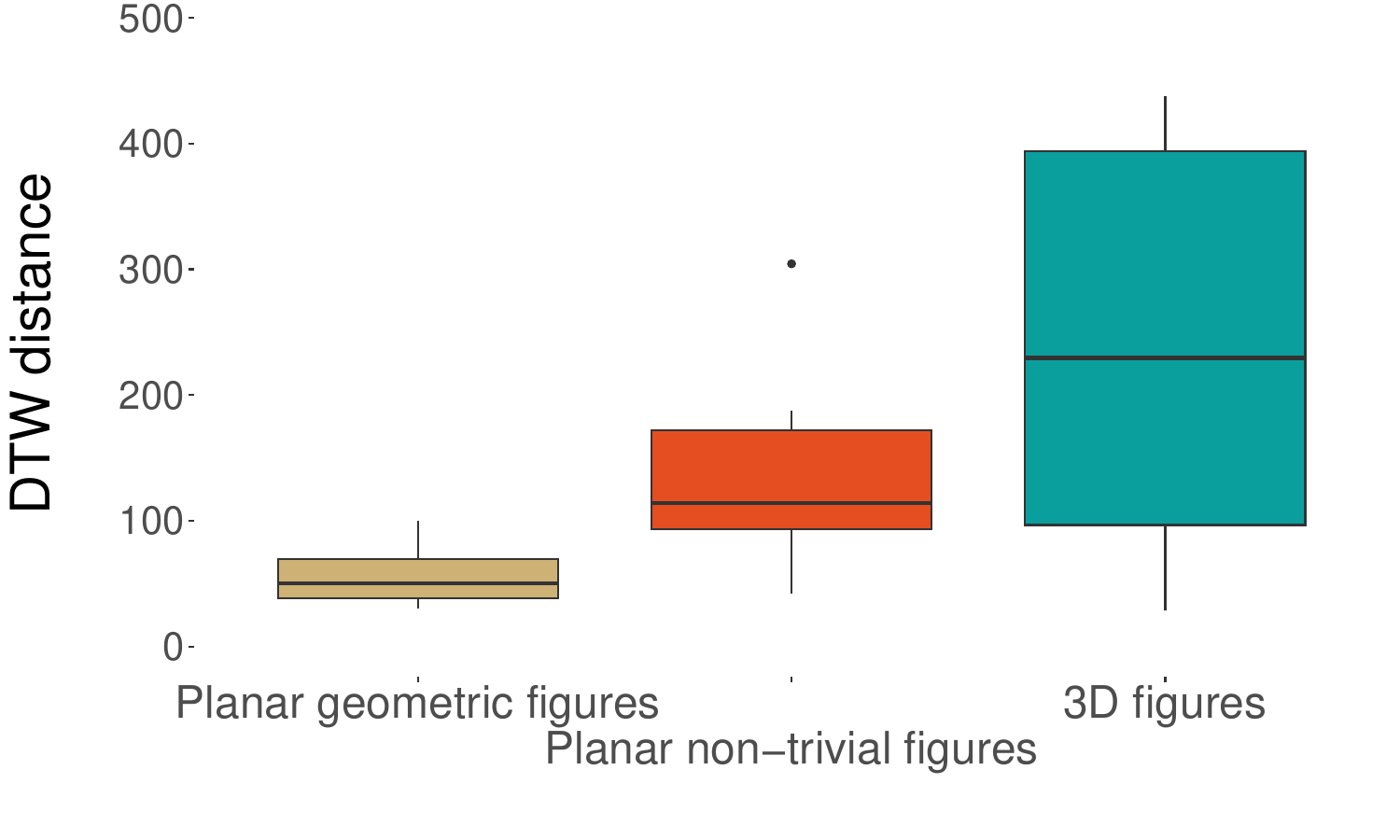}
	\end{minipage}
	\hfill
	\begin{minipage}{0.48\textwidth}
		\includegraphics[width=0.95\textwidth]{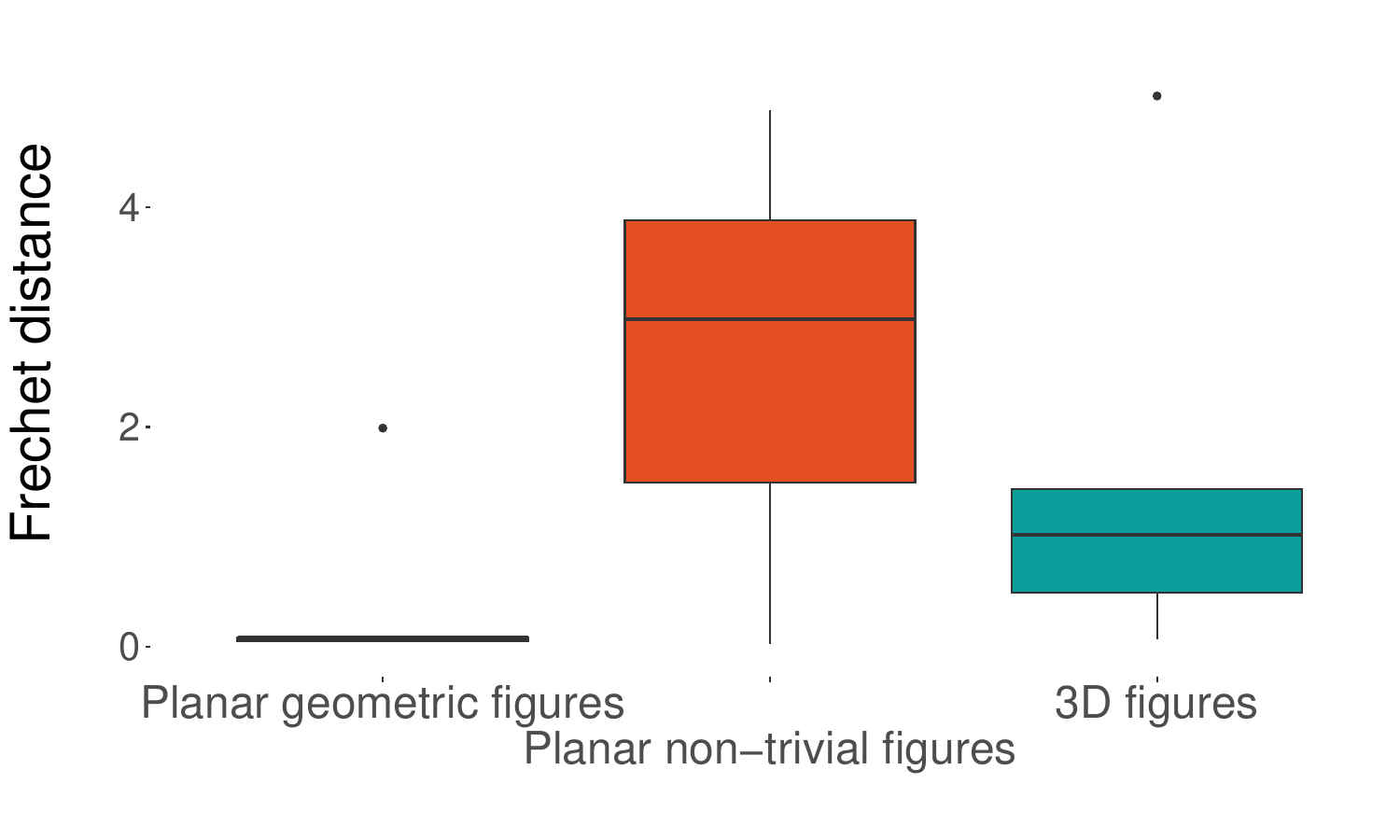}
		\includegraphics[width=0.95\textwidth]{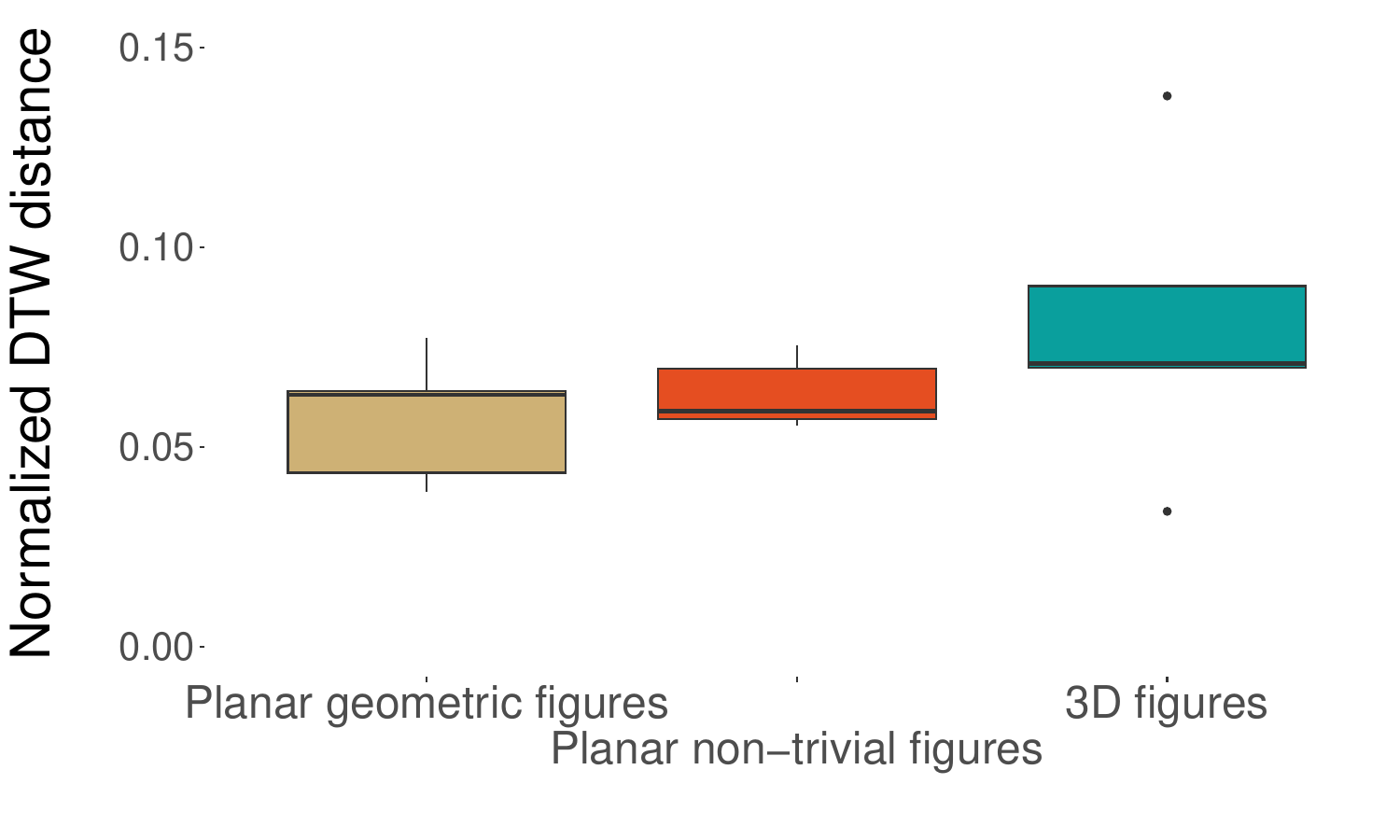}
	\end{minipage}
	\caption{Trajectory similarity of the demonstrated and programmed trajectory, distinguished by the scenario groups.}
	\label{fig:trajectory_similarities}
\end{figure}

As mentioned in \ref{ssec:experiment_setup}, we investigated whether the planning time saved comes at the expense of accuracy.
Therefore, we measured the Hausdorff, Frechet, Dynamic Time Warping and normalized Dynamic Time Warping distances between the 
programmed flight path and the trajectory based on the demonstration for each scenario. 
The results are presented in \ref{tab:similarities} as well as visualized in \ref{fig:trajectory_similarities}.

\newcommand{\lineimg}{
	\adjustbox{max width=2.5cm}{
		\tdplotsetmaincoords{0}{180}
		\begin{tikzpicture}[tdplot_main_coords, transform canvas={scale = 0.3, xshift = -0.5cm, yshift = 2cm}] 
			\begin{scope}[canvas is xy plane at z=1.0]
				\draw[line width=0.5mm] (2, 2, 0) -- (1, 2, 0);
			\end{scope}
		\end{tikzpicture}
	}
}

\newcommand{\squareimg}{
	\adjustbox{max width=2.5cm}{
		\tdplotsetmaincoords{0}{180}
		\begin{tikzpicture}[tdplot_main_coords, transform canvas={scale = 0.3, xshift = -0.5cm, yshift = 1.5cm}] 
			\begin{scope}[canvas is xy plane at z=1.0]
				\draw[line width=0.5mm] (2, 2, 0) -- (1, 2, 0) -- (1, 1, 0) -- (2, 1, 0) -- (2, 2, 0);
			\end{scope}
		\end{tikzpicture}
	}
}

\newcommand{\triangleimg}{
	\adjustbox{max width=2.5cm}{
		\tdplotsetmaincoords{0}{180}
		\begin{tikzpicture}[tdplot_main_coords, transform canvas={scale = 0.3, xshift = -0.5cm, yshift = 1.5cm}] 
			\begin{scope}[canvas is xy plane at z=1.0]
				\draw[line width=0.5mm] (2, 2, 0) -- (1, 2, 0) -- (1.5, 1.2, 0) -- (2, 2, 0);
			\end{scope}
		\end{tikzpicture}
	}
}

\newcommand{\circleimg}{
	\adjustbox{max width=2.5cm}{
		\tdplotsetmaincoords{0}{180}
		\begin{tikzpicture}[tdplot_main_coords, transform canvas={scale = 0.3, xshift = -1cm, yshift = 2.25cm}] 
			\begin{scope}[canvas is xy plane at z=1.0]
				\draw[line width=0.5mm, radius = 0.5] (1, 2, 0) circle;
			\end{scope}
		\end{tikzpicture}
	}
}

\newcommand{\ellipseimg}{
	\adjustbox{max width=2.5cm}{
		\tdplotsetmaincoords{0}{180}
		\begin{tikzpicture}[tdplot_main_coords, transform canvas={scale = 0.3, xshift = -1cm, yshift = 2.5cm}] 
			\begin{scope}[canvas is xy plane at z=1.0]
				\draw[line width=0.5mm] (1, 2, 0) ellipse (0.75 and 0.5);
			\end{scope}
			\end{tikzpicture}
		}
}

\newcommand{\obstacleone}{
	\adjustbox{max width=2.5cm}{
		\tdplotsetmaincoords{0}{180}
		\begin{tikzpicture}[tdplot_main_coords, transform canvas={scale = 0.25, xshift = -1cm, yshift = 2.5cm}] 
			\draw[-, line width=0.5mm, rounded corners=2pt] (0.5,2,1) -- (1,2,1);
			\draw[black, line width=0.5mm] (1, 2, 1) arc (180:360:0.5);
			\cuboid{1.5}{2}{1}{0.125}
			\draw[-, line width=0.5mm, rounded corners=2pt] (2,2,1) -- (2.5,2,1);
		\end{tikzpicture}
	}
}

\newcommand{\obstacletwo}{
	\adjustbox{max width=2.5cm}{
		\tdplotsetmaincoords{0}{180}
		\begin{tikzpicture}[tdplot_main_coords, transform canvas={scale = 0.25, xshift = -1cm, yshift = 2.5cm}] 
			\draw[-, line width=0.5mm, rounded corners=2pt] (0.5,2,1) -- (1,2,1);
			\draw[black, line width=0.85mm] (1, 2, 1) arc (180:360:0.5);
			\draw[black, line width=0.5mm] (2, 2, 1) arc (0:360:0.5);
			\cuboid{1.5}{2}{1}{0.125}
			\draw[-, line width=0.5mm, rounded corners=2pt] (2,2,1) -- (2.5,2,1);
		\end{tikzpicture}
	}
}

\newcommand{\slalom}{
	\adjustbox{max width=2.5cm}{
		\tdplotsetmaincoords{0}{180}
		\begin{tikzpicture}[tdplot_main_coords, transform canvas={scale = 0.25, xshift = -1cm, yshift = 2cm}] 
			\begin{scope}[canvas is xy plane at z=1.0]
				\draw[-, line width=0.5mm, rounded corners=2pt] (0.5,2,0) -- (1,2,0);
				\draw[line width=0.5mm] (1, 2, 0) arc (180:360:0.5);
				\draw[line width=0.5mm] (3, 2, 0) arc (0:180:0.5);
				\draw[line width=0.5mm] (3, 2, 0) arc (180:360:0.5);
				\draw[line width=0.5mm] (5, 2, 0) arc (0:180:0.5);
				\draw[-, line width=0.5mm, rounded corners=2pt] (5,2,0) -- (5.5,2,0);
			\end{scope}
			\cuboid{1.5}{2}{1}{0.125}
			\cuboid{2.5}{2}{1}{0.125}
			\cuboid{3.5}{2}{1}{0.125}
			\cuboid{4.5}{2}{1}{0.125}
		\end{tikzpicture}
	}
}

\newcommand{\virtualmaze}{
	\adjustbox{max width=2.5cm}{
		\tdplotsetmaincoords{0}{180}
		\begin{tikzpicture}[tdplot_main_coords, transform canvas={scale = 0.125, xshift = -2cm, yshift = 3.25cm}] 
			\begin{scope}[
				canvas is xy plane at z=1.0
				]
				\draw[line width=1mm, rounded corners=2pt, postaction={decorate}] (1, 1, 0) -- (4, 1, 0);
				\draw[line width=1mm, rounded corners=2pt, postaction={decorate}] (4, 1, 0) -- (4, 3.5, 0);
				\draw[line width=1mm, rounded corners=2pt, postaction={decorate}] (4, 3.5, 0) -- (1.5, 3.5, 0);
				\draw[line width=1mm, rounded corners=2pt, postaction={decorate}] (1.5, 3.5, 0) -- (1.5, 2, 0);
				\draw[line width=1mm, rounded corners=2pt, postaction={decorate}] (1.5, 2, 0) -- (3, 2, 0);
				\draw[line width=1mm, rounded corners=2pt, postaction={decorate}] (3, 2, 0) -- (3, 2.5, 0);
				\draw[line width=1mm, rounded corners=2pt, postaction={decorate}] (3, 2.5, 0) -- (2, 2.5, 0);
				\draw[line width=1mm, rounded corners=2pt, postaction={decorate}] (2, 2.5, 0) -- (2, 3, 0);
				\draw[line width=1mm, rounded corners=2pt, postaction={decorate}] (2, 3, 0) -- (3.5, 3, 0);
				\draw[line width=1mm, rounded corners=2pt, postaction={decorate}] (3.5, 3, 0) -- (3.5, 1.5, 0);
				\draw[line width=1mm, rounded corners=2pt, postaction={decorate}] (3.5, 1.5, 0) -- (1, 1.5, 0);
				\draw[line width=1mm, rounded corners=2pt, postaction={decorate}] (1, 1.5, 0) -- (1, 4, 0);
				\draw[line width=1mm, rounded corners=2pt, postaction={decorate}] (1, 4, 0) -- (4, 4, 0);
			\end{scope}
		\end{tikzpicture}
	}
}

\newcommand{\roominspection}{
	\adjustbox{max width=2.5cm}{
		\tdplotsetmaincoords{0}{180}
		\begin{tikzpicture}[tdplot_main_coords, transform canvas={scale = 0.25, xshift = -1cm, yshift = 2.3cm}] 
			\begin{scope}[canvas is xy plane at z=1.0]
				\draw[-, line width=0.5mm, rounded corners=2pt] (1,2,0) -- (2.5,2,0) -- (2.5,1,0) -- (2.5,2,0) -- (3.5,2,0) -- (3.5,3.5,0) -- (3.5,2,0) -- (4.5,2,0) -- (4.5,1.5,0);
				\draw[line width=0.5mm, radius=0.5] (4.5, 1, 0) circle;
			\end{scope}
		\end{tikzpicture}
	}
}

\newcommand{\house}{
	\adjustbox{max width=2.5cm}{
		\tdplotsetmaincoords{90}{0}
		\begin{tikzpicture}[tdplot_main_coords, transform canvas={scale = 0.35, xshift = -4.2cm, yshift = -1cm}] 
			\begin{scope}[canvas is xz plane at y=1.5]
				\draw[-, line width=0.4mm, rounded corners=2pt] (3,0.5) -- (2,0.5) -- (2,1.5) -- (2.5,2.366) -- (3,1.5) -- (2,1.5) -- (3,0.5) -- (3,1.5) -- (2,0.5);
			\end{scope}
		\end{tikzpicture}
	}
}

\newcommand{\cuboidimg}{
	\adjustbox{max width = 2.5cm}{
		\tdplotsetmaincoords{70}{160}
		\begin{tikzpicture}[tdplot_main_coords]
			\begin{scope}[
				transform canvas = {scale = 0.25}
				]
				\draw[line width=0.5mm, rounded corners=2pt, postaction={decorate}] (5, 3, 1) -- (2, 3, 1);
				\draw[line width=0.5mm, rounded corners=2pt, postaction={decorate}] (2, 3, 1) -- (2, 1, 1);
				\draw[line width=0.5mm, rounded corners=2pt, postaction={decorate}] (2, 1, 1) -- (2, 1, 2);
				\draw[line width=0.5mm, rounded corners=2pt, postaction={decorate}] (2, 1, 2) -- (2, 3, 2);
				\draw[line width=0.5mm, rounded corners=2pt, postaction={decorate}] (2, 3, 2) -- (5, 3, 2);
				\draw[line width=0.5mm, rounded corners=2pt, postaction={decorate}] (5, 3, 2) -- (5, 1, 2);
				\draw[line width=0.5mm, rounded corners=2pt, postaction={decorate}] (5, 1, 2) -- (2, 1, 2);
				\draw[line width=0.5mm, rounded corners=2pt, postaction={decorate}] (2, 1, 2) -- (2, 1, 1);
				\draw[line width=0.5mm, rounded corners=2pt, postaction={decorate}] (2, 1, 1) -- (5, 1, 1);
				\draw[line width=0.5mm, rounded corners=2pt, postaction={decorate}] (5, 1, 1) -- (5, 3, 1);
			\end{scope}
		\end{tikzpicture}
	}
}

\newcommand{\spiral}{ \adjustbox{max width = 2.5cm}{
		\tdplotsetmaincoords{60}{160}
		\begin{tikzpicture}[tdplot_main_coords, transform canvas={scale = 0.4}] 
			\draw[variable=\x,samples=100,domain=0:2*pi, smooth, very thick] plot ({0.5*sin(deg(\x))+2.5}, {0.5*cos(deg(\x))+1.5}, {0.5*(\x/(2*pi))^2+1});
			\draw[variable=\x,samples=100,domain=0:2*pi, smooth, very thick] plot ({0.5*sin(deg(\x))+2.5}, {0.5*cos(deg(\x))+1.5}, {0.5*(\x/(2*pi))^2+1.5});
		\end{tikzpicture}
	}
}

\newcommand{\staircase}{
	\adjustbox{max width=2.5cm}{
		\tdplotsetmaincoords{70}{150}
		\begin{tikzpicture}[tdplot_main_coords, transform canvas={scale = 0.4, xshift = -1cm}] 
			\draw[-, line width=0.5mm] (1,1,0.5) -- (2,1,1);
			\draw[line width=0.5mm] (2, 1, 1) arc (-90:90:0.25 and 0.5);
			\draw[-, line width=0.5mm] (2,2,1) -- (1,2,1.5);
			
		\end{tikzpicture}
	}
}

\newcommand{\slalomthreed}{
	\adjustbox{max width=2.5cm}{
		\tdplotsetmaincoords{50}{160}
		\begin{tikzpicture}[tdplot_main_coords, transform canvas={scale = 0.25, xshift = -1cm, yshift = 1.5cm}]

			\draw[-, line width=0.5mm, rounded corners=2pt] (0.5,2,1) -- (1,2,1);
			\begin{scope}[canvas is plane={O(1, 2, 1)x(2, 2, 1)y(1, 3, 0)}]
				\draw[black, line width=0.5mm] (0, 0, 0) arc (180:360:0.5);
				\draw[black, line width=0.5mm] (2, 0, 0) arc (0:180:0.5);
				\draw[black, line width=0.5mm] (2, 0, 0) arc (180:360:0.5);
				\draw[black, line width=0.5mm] (4, 0, 0) arc (0:180:0.5);
			\end{scope}
			\draw[-, line width=0.5mm, rounded corners=2pt] (5,2,1) -- (5.5,2,1);
			
			\cuboid{1.5}{1.5}{1}{0.125}
			\cuboid{2.5}{2.5}{1.25}{0.125}
			\cuboid{3.5}{1.5}{1}{0.125}
			\cuboid{4.5}{2.5}{1.25}{0.125}
		\end{tikzpicture}
	}
}

\newcommand{\roominspectiontwolevel}{
	\adjustbox{max width=2.5cm}{
		\tdplotsetmaincoords{50}{160}
		\begin{tikzpicture}[tdplot_main_coords, transform canvas={scale = 0.25, xshift = -1cm, yshift = 1.5cm}] 
			\begin{scope}[canvas is xy plane at z=1.0]
				\draw[-, line width=0.5mm, rounded corners=2pt] (1,2,0) -- (2.5,2,0) -- (2.5,1,0) -- (2.5,2,0) -- (3.5,2,0) -- (3.5,3.5,0) -- (3.5,2,0) -- (4.5,2,0) -- (4.5,1.5,0);
				\draw[line width=0.5mm, radius=0.5] (4.5, 1, 0) circle;
			\end{scope}
			\draw[-, line width=0.5mm, rounded corners=2pt, postaction={decorate}] (1,2,1) -- (1,2,1.5);
			\begin{scope}[canvas is xy plane at z=1.5]
				\draw[black, -, line width=0.5mm, rounded corners=2pt] (1,2,0) -- (2,2,0) -- (2,3.5,0) -- (2,2,0) -- (3,2,0) -- (3,2.5,0);
				\draw[black, line width=0.5mm] (3, 3, 0) ellipse (0.25 and 0.5);
			\end{scope}
		\end{tikzpicture}
	}
}

\begin{table}[htb]
	\centering
	\caption{Trajectory Similarity Measures for all scenarios}
	\label{tab:similarities}
	\renewcommand{\arraystretch}{1.25}
	\begin{tabular}{p{4.3cm}>{\raggedleft}m{2.2cm}cccc}
		\toprule
		Scenario Name                 &                         & Hausdorff & Frechet &   DTW   & avg. DTW \\ \midrule
		Line                          & \lineimg                &   0.076   &  1.989  & 30.525  &  0.077   \\
		Square                        & \squareimg              &   0.070   &  0.036  & 69.339  &  0.044   \\
		Triangle                      & \triangleimg            &   0.075   &  0.052  & 38.424  &  0.063   \\
		Circle                        & \circleimg              &   0.075   &  0.081  & 50.056  &  0.039   \\
		Ellipse                       & \ellipseimg             &   0.085   &  0.084  & 100.091 &  0.064   \\ \midrule
		Fly around obstacle (1)       & \obstacleone            &   0.090   &  2.969  & 41.854  &  0.055   \\
		Fly around obstacle (2)       & \obstacletwo            &   0.093   &  2.998  & 102.798 &  0.075   \\
		Virtual slalom                & \slalom                 &   0.120   &  4.888  & 90.120  &  0.059   \\
		Virtual maze                  & \virtualmaze            &   0.112   &  4.175  & 304.325 &  0.059   \\
		Room inspection               & \roominspection         &   0.140   &  0.019  & 187.737 &  0.056   \\
		The house of Wolfgang Reif    & \house                  &   0.138   &  1.000  & 125.505 &  0.073   \\ \midrule
		Cuboid                        & \cuboidimg              &   0.206   &  0.064  & 393.892 &  0.090   \\
		Spiral                        & \spiral                 &   0.135   &  1.017  & 96.676  &  0.071   \\
		Staircase                     & \staircase              &   0.078   &  1.434  & 29.034  &  0.034   \\
		3D slalom                     & \slalomthreed           &   0.208   &  5.015  & 229.538 &  0.138   \\
		Room inspection on two levels & \roominspectiontwolevel &   0.195   &  0.489  & 437.796 &  0.070   \\ \bottomrule
	\end{tabular}
\end{table}

\ref{fig:trajectory_similarities} clearly illustrates, that the more complex the scenarios become (categories \textit{Planar non-trivial figures} and \textit{3D figures}), the greater
the distances are, i. e. the trajectory based on the demonstration is less similar to the programmed trajectory.
As the Dynamic Time Warping distance depends on the trajectory length, the increasing distance for more complex scenarios is not surprising.
The average Frechet distance is 1.644, but there are strong outliers, as evidenced by a standard deviation of 1.813.

In contrast, the average of all Hausdorff distances is with 0.1186 more meaningful, as the standard deviation is relatively lower with 0.048. This average distance can be interpreted as, 
on average over all scenarios, a maximum distance from any point of the demonstrated trajectory to the programmed trajectory of about 11.8 cm.
Furthermore, also averaged over all scenarios, the average distance between the demonstrated, optimally matched, and the related programmed trajectory is less than 6.67 cm (SD = 2.4) - 
called the normalized DTW distance. 

In general, the results indicate that the \textit{Drawjectory} workflow accelerates the trajectory planning process ($-78.7$ seconds on average).
In particular, when planning more complex trajectories, the programming of these trajectories requires a more time (+$82.2$ seconds), while the demonstration
exhibits a minimal additional overhead (+$21.6$ seconds).
This acceleration comes at the expense of accuracy, as shown by the Hausdorff or normalized DTW distance increasing by 0.0876 and 0.0232 (in m), respectively, for longer and more complex scenarios.

To conclude, the \textit{Drawjectory} workflow is a fast and intuitive approach to manual trajectory planning 
while maintaining a high level of precision. In particular, when used in conjunction with the \textit{Hybrid Editor}, 
so programmed trajectories, it can be an efficient and effective method for manual trajectory planning.

\subsection{Threats to Validity}
\label{ssec:validity_threats}

As this study is an experiment, it is subject to a number of potential threats.
Accordingly, we discuss these threats to \textit{internal}, \textit{external}, \textit{construct} and the \textit{conclusion} validity \cite{Wohlin2012Experimentationsoftwareengineering}
and how we addressed them.

\paragraph{Internal validity}

One potential threat to the internal validity is the different time points at which the two input modalities were tested. To address this issue, both input
modalities were tested without any time pressure, starting at the same time of day. Additionally, knowledge of the procedure could have a negative effect on the internal validity. 
However, as the test person knew all the scenarios exactly, had a lot of experience with the programs and had carried out a test run on two sample scenarios, 
there was no familiarity effect after testing the first input modality.
In general, using the same scenarios and the same version of the programs (\textit{Hybrid Editor} and \textit{Drawjectory Control Panel}) 
increased the internal validity.

\paragraph{External validity}

The external validity is mainly threatened by the fact that the scenarios described in \ref{ssec:experiment_setup} were rather simple and are not necessarily comparable
to potential real-world scenarios, but they do contain many base cases from which real-world scenarios could be built.
Further factors threatening the external validity are:
\begin{enumerate}
	\item The used tracking system in the \textit{Quadcopter Lab}, as the \textit{OptiTrack}-system cannot be applied in every situation, especially outside laboratories.
	\item The experiment was conducted by a person experienced in both demonstrating and programming trajectories.
\end{enumerate}
Although the \textit{OptiTrack}-system may not be widely used in the private sector, it is used in industry \cite{Optitrack} and could possibly be replaced by another, more accessible
tracking system as the entire workflow does not require a specific tracking system.

In general, the \textit{Drawjectory} workflow on its own is only a concept, which can be implemented by any person in any situation
and is therefore generalizable.

\paragraph{Construct validity}

To increase the construct validity we employed different measures for the trajectory similarity, which allowed us to cross-check the results.

\paragraph{Conclusion validity}

The employed measures for assessing the trajectory similarity between the demonstrated and the programmed trajectories are reliable and widely used in the literature \cite{Chang2024Trajectorysimilaritymeasurement,Toohey2015Trajectorysimilaritymeasures}.
Furthermore, the planning time was objectively measured, as it was automatically determined. As the experiment was conducted in the quadrocopter laboratory by one author, the results were not affected by any external disturbances.

%% file: sections/related_work.tex
\section{Related Work}
\label{sec:related_work}

The programming of robots is a mature field of research, particularly in relation to industrial robots\cite{Heimann2020}. The difference to the drones considered here is that industrial robots typically have fixed axes and can therefore be positioned very precisely. Drones, on the other hand, hover in the air and therefore cannot actually be held in exact positions. Hence, in the following, we focus on related work that deals with the trajectory planning of drones according to the approaches used in this work.

In \textquote{A Hybrid Editor for Fast Robot Mission Prototyping}\cite{Witte2019HybridEditorFast}, Witte et al. introduce a high-level domain-specific language (DSL) that can be used to directly set waypoints the drone has to pass by. The programming is supported by a \textit{RQt} plugin including a visualization of the calculated trajectory and the functionality that any move of the waypoints in the visualization is reflected in the source code.
The authors claim that this editor allows novice and advanced programmers to quickly prototype flight routes because of visual feedback, but do not provide a user study. For this paper, we extended this language with additional constructs as described in \ref{sec:quadcopter_lab} and evaluated how fast a trajectory can be created by using either the DSL or the demonstration approach. 

Hoppe et al. \cite{Hoppe2019Dronosflexibleopen} propose a framework for rapid prototyping of drone routines called \textit{DronOS}. Similarly to our approach, the authors assess \textit{DronOS} by using three different methods for planning a flight route for a quadrocopter:

\begin{itemize}
    \item \textit{Unity Scripting}: the flight route is planned using waypoints in a 3D environment using drag-and-drop. This process is analogous to the interactive waypoints presented in \cite{Witte2019HybridEditorFast}.
    \item \textit{Vive Scripting}: Waypoints are set using a controller in the real world and can be edited later in a virtual environment. 
    \item \textit{Vive Realtime}: The user holds a controller and points to different targets. The quadrocopter follows these targets in realtime and records the used flight path. 
\end{itemize}

The difference between the \textit{Vive Realtime} method and our approach presented here is that we demonstrate the complete path and not just individual points. In Drawjectory, the path can be drawn without a drone having to fly it in realtime. However, the challenge in our approach is to automatically extract appropriate waypoints from the drawn trajectory.
The presented methods were evaluated in a small user study with 12 participants, whose task completion time as well as the mental workload using a NASA-TLX questionnaire \cite{Hart1988DevelopmentNASATLX} were measured. Similar to our results, they conclude that real-time demonstration was the fastest method, while \textit{Unity scripting} is the most accurate and \textit{Vive scripting} is the most intuitive.

Yau et al. \cite{Yau2020Howsubtlecan} try to improve flight route planning by introducing interaction models for controlling quadrocopters with one hand. Their challenge was to map the six degrees of freedom of a flying drone to the limited options of one hand controllers. They evaluated four different mappings by a user study with 32 participants with respect to the task completion time and mental load (NASA-TLX) for very simple task. Based on these results, the conducted a second user study for the best option for more complex tasks. The participants ``found the device easy to use and easy to learn''\cite{Yau2020Howsubtlecan}. 

These proposed input modalities concentrate on how to achieve the most practical quadrocopter live control method, rather than focusing on a precise and intuitive (pre-)planning of a trajectory, which is the main goal of our approach. 
In the following three programming by demonstration approaches, trajectories are not planned for drones but for classic robots. What all these approaches have in common is the handling of inaccuracies when demonstrating a trajectory, but with different techniques. 

Melchior et al. \cite{Melchior2012Graphbasedtrajectory} use a neighbor graph to identify correspondences between multiple imperfect demonstrations. Compared to Drawjectory, this process has the weakness that it takes several demonstrations to produce a trajectory close to the desired one. 

Aleotti et al. \cite{Aleotti2005Trajectoryreconstructionnurbs} propose a workflow for learning robot trajectories based on single or multiple demonstrations. These demonstrations are tracked by a \textit{CyberTouch VR glove}, which returns a set of points in three-dimensional space. Similar to \textit{Drawjectory}, the user can change the position of selected waypoints at the end. The difference, however, is that we are not drawing the trajectory in a virtual space but in real space, which makes the input more intuitive.   

The third presented work by Hwang et al. \cite{Hwang2003Mobilerobotsyour} is in several aspects closer to our approach: Firstly, the user draws the desired path with a finger or a stylus on a touchscreen. Then, an improved version of a corner detection algorithm extracts significant waypoints, which can be edited afterwards. Finally, a smooth trajectory is interpolated by using so-called piece-wise cubic Bezier curves (PCBC). 
The main drawback of this method is that it is designed for two-dimensional space
and the PCBCs, in contrast to the natural cubic splines employed in our implementation, are not twice continuously differentiable at the waypoints \cite{Hwang2003Mobilerobotsyour,Olkin2021AutonomousQuadrotorTrajectory}. 
However, this is necessary to ensure a smooth trajectory, which is the goal of the trajectory planning in the \textit{Drawjectory} workflow.

%% file: sections/conclusion.tex
\section{Conclusion}%
\label{sec:conclusion}

Quadcopters are becoming an increasingly common part of our everyday lives, 
often controlled automatically thanks to ever-improving algorithms for automatic planning. In some scenarios, however, planning
quadrocopter flight paths must still be done manually.
Therefore, there arises a need for an intuitive but still accurate process to manually plan flight paths.
To solve this problem, we propose the fast and intuitive \textit{Drawjectory} workflow, which allows \textquote{programming} flight paths for a quadrocopter by demonstration.

The workflow consists of three main steps: i) demonstrating the flight path by tracking a gesture wand, %
ii) selecting points from this flight path and planning a trajectory based on these points, %
and iii) optionally editing the planned trajectory with input from the user. %

The presentation of this workflow is supported by a proof-of-concept implementation in ROS2 that runs in the \textit{Quadcopter Lab} of the University of Ulm. 
Based on a demonstration using motion capture, the implementation plans a trajectory from selected waypoints using a natural cubic spline interpolation.
Both the demonstrated flight path and the planned trajectory are visualized in the \emph{RViz} tool and can optionally be edited by shifting, moving, or scaling waypoints in the \texttt{Drawjectory Control Panel}.

Finally, we evaluate the \textit{Drawjectory} workflow and its implementation.
We find that while maintaining high precision the \textit{Drawjectory} workflow intuitively accelerates the trajectory planning process greatly compared to programming, 
More specifically, the proposed workflow is, on average, 78.7 seconds faster than programming and excels especially when planning complex trajectories, 
but causes an average deviation of approximately 6.67 cm and an average maximum point-to-trajectory deviation of 11.9 cm.

However, during the development of the workflow and its proof-of-concept implementation, several issues arose that could not be solved within the scope of this work:

Recalling the goals of the workflow, in particular the requirement that this workflow should use only few technical aids, not all goals are achieved, due to the fact that the motion capture system is immobile and difficult to set up.
This limitation could potentially be overcome by instead using VR devices requiring much less setup.

Furthermore, in this work, the orientation of the quadrocopter is not considered at all, as it is not necessary for flying along a path. However, in real world scenarios, for example when a camera is mounted on the quadrocopter, the orientation is important.
The tracking and trajectory planning process could be adapted accordingly in future work.

Moreover, the waypoint selection strategy and the interpolation method for the trajectory are quite simple.
It could be interesting to investigate how different strategies affect --- for example --- the accuracy in respect to the user's intentions.

Lastly, one potential follow-up work could investigate how the intention of the demonstration can be identified automatically.
The user could then be suggested corrections based on the predicted intention.